\documentclass[ aip, amsmath, amssymb, reprint,]{revtex4-1}

\usepackage{graphicx}
\usepackage{dcolumn}
\usepackage{bm}

\newcommand*{\mat}[1]{\boldsymbol{#1}}

\newcommand*{\mA}{{\mat{A}}}

\newcommand*{\mC}{{\mat{C}}}

\newcommand*{\mF}{{\mat{F}}}

\newcommand*{\mH}{{\mat{H}}}

\newcommand*{\mS}{{\mat{S}}}
\newcommand*{\mT}{{\mat{T}}}
\newcommand*{\mU}{{\mat{U}}}
\newcommand*{\mV}{{\mat{V}}}

\newcommand*{\vb}{\boldsymbol{b}}
\newcommand*{\vc}{\boldsymbol{c}}

\newcommand*{\vg}{\boldsymbol{g}}

\newcommand*{\vp}{\boldsymbol{p}}

\newcommand*{\vv}{\boldsymbol{v}}

\def\bea{\begin{eqnarray}}
	\def\eea{\end{eqnarray}}
\def\ben{\begin{equation}}
	\def\een{\end{equation}}
\def\benu{\begin{enumerate}}
	\def\enu{\end{enumerate}}

\def\lsim {\ifmmode {\buildrel<\over\sim}}

\def\sss{\scriptscriptstyle\rm}

\def\1var{(\bx_1...\bx\N)}

\def\br{{\bf r}}

\def\b1{{\bf 1}}
\def\bx{{x}}

\def\xc{_{\sss XC}}
\def\KS{_{\sss KS}}

\def\N{_{\sss N}}
\def\H{_{\sss H}}

\def\ext{_{\rm ext}}

\draft 

\begin{document}
	
	\title{Inverse Kohn-Sham Density Functional Theory: Progress and Challenges} 
	
	\author{Yuming Shi}
	\affiliation{Department of Physics and Astronomy, Purdue University, West Lafayette, Indiana 47907}
	
	\author{Adam Wasserman}
	\email[]{awasser@purdue.edu}
	\affiliation{Department of Physics and Astronomy, Purdue University, West Lafayette, Indiana 47907}
	\affiliation{Department of Chemistry, Purdue University, West Lafayette, Indiana 47907}
	\affiliation{Purdue Quantum Science and Engineering Institute, Purdue University}
	
	\date{\today}
	
	\begin{abstract}
		Inverse Kohn-Sham (iKS) problems are needed to fully understand  the one-to-one mapping between densities and potentials on which Density Functional Theory is based. They are also important to advance computational schemes that rely on density-to-potential inversions such as the Optimized Effective Potential method and various techniques for density-based embedding. Unlike the forward Kohn-Sham problems, numerical iKS problems are ill-posed and can be unstable. We discuss some of the fundamental and practical difficulties of iKS problems with constrained-optimization methods on finite basis sets. Various factors that affect the performance are systematically compared and discussed, both analytically and numerically, with a focus on two of the most practical methods: the Wu-Yang method (WY) and partial-differential-equation constrained-optimization (PDE-CO).  Our analysis of the WY and PDE-CO highlights the limitation of finite basis sets and the importance of regularization. We introduce two new ideas that will hopefully contribute to making iKS problems more tractable: (1) A correction to the WY method that utilizes the null space of the relevant Hessian matrices; and (2) A finite potential basis-set implementation of the PDE-CO method. We provide an overall strategy for performing numerical density-to-potential inversions that can be directly adopted in practice.  We also provide an Appendix with several examples that can be used for benchmarking. 
	\end{abstract}
	
	\maketitle 
	
	\section{Introduction}
	{\bf In principle {\em vs.} in practice:} Kohn-Sham (KS) density functional theory (DFT) \cite{hohenberg1964inhomogeneous, kohn1965self} has long been the most widely used method for electronic-structure calculations in condensed matter physics and quantum chemistry computations \cite{burke2012perspective,verma2020status}. 
	KS-DFT is formally exact in the sense that, given the exact exchange-correlation (XC) energy functional, a numerically exact solution of the self-consistent KS-DFT equations is guaranteed to yield the exact ground-state density $n(\br)$ and energy for any system of $N$ electrons in a time-independent external potential $v(\br)$. The word "exact", which has been used 4 times already in this introductory paragraph, is sometimes dismissed with scorn when confronted with results in practice. The results of actual DFT calculations are evidently not exact. Nevertheless, it is the proven existence of an {\em exact} one-to-one correspondence between ground-state densities and potentials which has given impetus to the development of approximations in DFT.  A different one-to-one mapping exists for any choice of electron-electron interaction.  Nature's choice is the Coulomb interaction, but the correspondence can be established also for a fictitious system of non-interacting electrons. Calculations are simpler in this fictitious world, and the one-to-one maps allow one to connect the answers back to the real world, {\em in principle}.  In practice, the exchange-correlation energy $E\xc[n]$, ``nature's glue" \cite{kurth2000role}, needs to be approximated. 
	This paper discusses another clash between ``in-principle" and ``in practice": The density-to-potential mappings are in principle one-to-one. For a given density $n(\br)$ and choice of approximate $E\xc[n]$, there is in principle only one XC potential $v\xc(\br)$ corresponding to $n(\br)$. That potential is the functional derivative of $E\xc[n]$ with respect to the density, evaluated at that density. The potential $v\xc(\br)$ should be calculable from the given density by solving an inverse problem, in principle. In practice, unfortunately, the process is a numerical minefield. 
	For a user's choice of approximate $E\xc[n]$, most quantum-chemistry codes solve the {\em forward} KS-DFT problem, i.e. self-consistently calculate $n(\br)$ and, from it, the total energy for a given $v(\br)$ and $N$. Another output of this calculation is the self-consistent $v\xc(\br)$. Much less common are codes that can solve the inverse Kohn-Sham problem (iKS) to find the XC potential corresponding to a given density\cite{nam2020kohn, unsleber2018serenity}. But why would one want to do such calculations?  We list four answers below:
	
	\begin{itemize}
		\item {Exploring the Hohenberg-Kohn \cite{hohenberg1964inhomogeneous} and Runge-Gross \cite{runge1984density} mappings}: More generally, density-to-potential inversions are useful for calculating numerically exact XC potentials from numerically exact densities, i.e. a point-by-point exploration of the Hohenberg-Kohn and Runge-Gross mappings. In ground-state DFT, inversions can reveal features of XC potentials that encode the physics of strong correlations \cite{HTR09,hodgson2017interatomic} and are missed by approximate functionals. In time-dependent DFT, inversions can also shed light on properties of the exact time-dependent XC potentials needed to describe electronic dynamics 
		\cite{TK09,FNRM16}, and to propose approximations that go beyond the adiabatic limit \cite{MZCB04, M16, ramsden2012exact, mancini2014adiabatic}, as needed for describing multiple and charge-transfer excitations in molecules or excitonic transitions in solids.
		\item {Analyzing errors of approximate XC functionals}: iKS can be utilized for measuring density-driven errors in KS-DFT \cite{nam2020measuring};
		iKS from numerically exact densities can assist on the development of explicit XC functionals. 
		There are also recent examples of the use of iKS to improve XC functionals in the field of nuclear DFT \cite{naito2019improvement, accorto2020first, accorto2021nuclear}. Furthermore, machine-learning approaches under development \cite{brockherde2017bypassing, nagai2018neural, kalita2021learning, manzhos2020machine} also take inverted XC potentials as input.  The efficiency and accuracy of the iKS methods
		is critically important here.
		
		\item {Quantum Embedding}: Various modern variants of density-based subsystem and embedding approaches \cite{JN14,WSZ15} employ potential reconstruction techniques \cite{GAMM10,NWW11,HPC11}. 
		Here, typically, a target density is updated many times in a self-consistent loop and the potential corresponding to that target density must be found at each step of the loop, implying a significant additional computational cost to calculations that do not make use of efficient inversion techniques. 
		\item {Optimized Effective Potentials} (OEP): Approximate expressions for $E\xc[n]$ can be either explicit or implicit functionals of the density.  Explicit functionals are usually preferred for their simplicity and efficiency. However, implicit XC functionals \cite{KK08} that depend on the KS orbitals are often used when higher accuracy is needed \cite{levy1979universal, perdew2005prescription}. In particular, OEP methods are being increasingly used for a variety of applications \cite{JYZP17,JZCS17,FSRA18}. 
		The OEP method is often the bottleneck of the calculations. The direct OEP method developed by \citeauthor{yang2002direct}\cite{yang2002direct} shares many of the same features of iKS discussed here \cite{gidopoulos2012nonanalyticity, bulat2007optimized, heaton2008optimized}. Improvements on inversion methods would potentially offer significant speed-ups allowing for more widespread adoption of OEP calculations. 
	\end{itemize}
	
	{\bf iKS methods:} Several methods exist for solving the iKS problem through self-consistent density-based calculations \cite{aryasetiawan1988effective, gorling1992kohn, zhao1992quantities, zhao1993constrained, wang1993construction, peirs2003algorithm, kadantsev2004variational, wagner2014kohn, ou2018potential}. The connections between many of these methods have been cogently explored recently by Kumar et al.\cite{kumar2019universal, kumar2020general}. An entirely different approach that makes use of the wavefunction (instead of the density) has been developed by \citeauthor{ryabinkin2015reduction} 
	(mRKS) and shown to provide arguably the best inversion results to date \cite{ryabinkin2015reduction, ospadov2017improved}. mRKS uses as input the one- and two-particle reduced density matrices. Its computational cost is thus out of reach for calculations on large systems. A detailed discussion explaining the success of wavefunction based methods compared to pure density inversions was provided recently by Kumar et al.\cite{kumar2020accurate}. We
	concentrate on {\em pure Kohn-Sham inversions}, referred to as iKS in this Perspective. We use mRKS in this article as benchmark to compare with other purely density-based approaches. Methods other than self-consistent calculations have also been designed, many of which feature constrained optimizations \cite{zhao1994electron, wu2003density, nafziger2017accurate, jensen2018numerical, kanungo2019exact, kumar2020general, callow2020density}.\par
	
	In spite of all the methods in hand, iKS problems are still difficult. First, different methods feature different capabilities regarding accuracy and efficiency. Many of them have been only tested on single atoms or diatomic molecules for illustration. Some accurate methods, such as mRKS, are difficult to apply to molecules with more than $\sim$10 atoms.
	\citeauthor{kanungo2019exact} recently implemented the PDE-Constrained Optimization method (PDE-CO) on a systematically
	improvable finite-element basis that provided results for polyatomic systems with an accuracy comparable to that of mRKS using affordable computation resources \cite{kanungo2019exact}. KS inversions on finite basis sets are generally considered as ill-posed. A problem is well-posed if a solution exists, it is unique and continuously changes with the input as defined by \citeauthor{hadamard1902problemes} \cite{hadamard1902problemes}. If a problem is not well-posed, it is then ill-posed. Analogous inverse problems in many fields are generally known for their instabilities. In the case of iKS, the uniqueness is guaranteed by the Hohenberg-Kohn theorem. The existence is known as the $v$-representability problem. In discretized systems, densities are ensemble $v$-representable \cite{chayes1985density, ullrich2002degeneracy, lammert2006coarse}. The existence is usually assumed to be true. The main problem of KS inversion lays in the continuity of the density-to-potential mapping \cite{jensen2018numerical}. Since the kinetic operator in the KS equation plays a role of regulator, different XC potentials can reproduce very similar densities \cite{nagai2018neural, li2020kohn, nagai2020completing}. In practical KS inversion calculations on finite basis sets, many factors can lead to unphysical oscillations/overfitting to the final $v\xc(\br)$. Therefore, special consideration and regularization is often essential for reasonable results. Simple tricks can often greatly improve an inversion when using one particular inversion method but those same tricks may be totally unhelpful when used in combination with other methods. Many of these tricks depend on error cancellations of some form \cite{gaiduk2013removal, kanungo2019exact, wu2003direct, ospadov2017improved, bulat2007optimized, heaton2008optimized, jacob2011unambiguous, nam2020kohn}, making it extremely difficult to predict when they will work or fail.
	
	{\bf Organization of this paper:} 
	We focus on two of the most efficient constrained-optimization methods for iKS,  the Wu-Yang method \cite{wu2003direct} (WY) and PDE-Constrained Optimization \cite{nafziger2017accurate, jensen2018numerical} (PDE-CO). First, we review the theory behind both methods and their implementation on finite basis sets, including the first implementation of PDE-CO on Gaussian potential basis sets.
	Except for the well-known drawbacks of the finite potential basis set (PBS), PBS significantly improves the efficiency and helps to control problems that one would encounter in a general PDE-CO problem \cite{jensen2018numerical, kanungo2019exact, vogel2002computational}.
	Different factors that influence the stability of the inversion are systemically discussed and compared, both analytically and numerically, including finite basis sets, regularization/corrections, optimization methods and guide potentials. 
	There is no way of avoiding the use of many different acronyms for the methods, algorithms, and basis sets employed here. Table \ref{list:acronyms} compiles the acronyms we use most.  We highlight our use of ``CX/CY" when cc-pCVXZ is being used as the basis set to expand the orbitals and cc-pCVYZ as the basis set to expand the potentials. 
	The input density is CCSD unless further specified. The exact XC potential data comes from quantum Monte Carlo calculations \cite{umrigar1994accurate, filippi1996recent}. All the calculations are implemented on Psi4 \cite{parrish2017psi4, smith2018psi4numpy}. 
	Atomic units are used throughout.
	\begin{table}[h!]
		\centering
		\begin{tabular}{|l l|}
			\hline
			CCSD& coupled-cluster singles-\\
			& and-doubles\\
			\hline
			CX  & cc-pCVXZ\\
			\hline
			CX/CY & The OBS/PBS is CX/CY\\
			\hline
			FA & Fermi-Amaldi\\
			\hline
			iKS & inverse Kohn-Sham\\
			\hline
			KS  & Kohn-Sham\\
			\hline
			LDA & Local Density Approximation \\
			\hline
			mRKS & modified Ryabinkin–\\
			&Kohut–Staroverov\\
			\hline
			NSC & null-space correction\\
			\hline
			OBS & atomic orbital basis set\\
			\hline
			NSC & null space correction\\
			\hline
			PBS & potential basis set\\
			\hline
			PBE & Perdew-Burke-Ernzerhof\\
			\hline
			PDE & partial differential equation\\
			\hline
			PDE-CO &  PDE-Constrained optimization\\
			\hline
			QMC & quantum Monte Carlo\\
			\hline
			TSVD & truncated singular value \\
			& decomposition\\
			\hline
			WY  & the Wu-Yang method\\
			\hline
			XC  & exchange-correlation\\
			\hline
		\end{tabular}
		\caption{Acronym List}
		\label{list:acronyms}
	\end{table}

	\section{Theory}
	\subsection{The Wu-Yang Method (WY)} 
	The central idea of the WY method follows directly the Levy
	constrained-search approach for the KS kinetic energy functional \cite{levy1979universal, kohn1965self}: optimizing the non-interacting kinetic energy under the constraint that the density matches a target. Unlike many other methods that depend on some sort of self-consistent calculation, the WY method relies on gradient and Hessian-based optimizations. Thus the WY method can be very easily implemented with a standard general optimizer and is robust for most systems. On the other hand, given the ill-posed nature of iKS and the ill-conditioning of the Hessian matrices, there can be numerical problems and regularization is usually essential.\par The Lagrangian for the WY constrained optimization is
	\begin{equation}\label{equ:WuYangL}
		\begin{split}
			&W[\Psi_{\rm det}[v\KS], v\KS]\\
			= &T_s[\Psi_{\rm det}] + \int d\br v\KS(\br)\{n(\br)-n_{\rm in}(\br)\},
		\end{split}
	\end{equation}
	where $n_{\rm in}(\br)$ is the target input density and $n(\br)$ is:
	\begin{subequations}
		\label{Density}
		\begin{align}
			n(\br) &= N\int d\br_2\dots d\br_N |\Psi_{\rm det}(\br,\br_2,\dots,\br_N)|^2 \label{Density1}\\
			&= 2\sum_i^{N/2} |\psi_i(\br)|^2\label{Density2}.
		\end{align}
	\end{subequations}
	The $\Psi_{\rm det}$ in Eqs.(\ref{equ:WuYangL})-(\ref{Density}) is the KS Slater determinant consisting of $N/2$ doubly occupied orthonormal KS orbitals $\{\psi_i\}$. Our discussions will be limited to spin-unpolarized systems for convenience. $v\KS(\br)$ in Eq.(\ref{equ:WuYangL}) is the KS potential:
	\begin{equation}\label{eqn:eigen}
		[\hat{T}+v\KS(\br)]\psi_i(\br)=\epsilon_i\psi_i(\br)~~,
	\end{equation}
	but appears in Eq.(\ref{equ:WuYangL}) as a Lagrange multiplier because of the necessary condition on the stationary point:
	\begin{equation}\label{dWdphi}
		\frac{\delta W[\Psi_{\rm det}, v\KS]}{\delta n(\br)} = 0.
	\end{equation}
	Here, a Lagrangian dual problem of the Zhao-Morrison-Parr \cite{zhao1994electron} problem is solved and $v\xc(\br)$ is the dual variable of the density \cite{wu2014variational, kumar2020general}. Therefore, (\ref{equ:WuYangL}) needs to be maximized. This will also be proven later through features of the hessian matrices.\par
	From (\ref{equ:WuYangL}):
	\begin{equation}\label{equ:WYGrad}
		\frac{\delta W[\Psi_{\rm det}[v\KS], v\KS]}{\delta v\KS(\br)} = n(\br)-n_{\rm in}(\br)
	\end{equation}
	and
	\begin{equation}\label{equ:WYHess}
		\begin{split}
			&\frac{\delta^2 W[\Psi_{\rm det}[v_{\rm KS}], v_{\rm KS}]}{\delta v\KS(\br)\delta v\KS(\br')}=\frac{\delta n(\br)}{\delta v_{KS}(\br')}\\
			=& 2\sum_i^{occ.}\sum_a^{unocc.}
			\frac{\psi_i^*(\br)\psi_a(\br)\psi_i(\br')\psi_a^*(\br')}{\epsilon_i - \epsilon_a}.
		\end{split}
	\end{equation}
	\par
	\subsection{PDE-Constrained Optimization (PDE-CO)} 
	Based on the Hohenberg-Kohn theorem \cite{hohenberg1964inhomogeneous} and assuming non-degeneracy, one would expect from an ``exact" input density the exact KS potential. The continuity condition in the definition of well-posed problems is not proven, but it is generally assumed to hold (at least around the "exact" density): a more accurate potential is expected to yield a more accurate density and vice versa. To find the best $v\KS(\br)$ given the limitation of a basis set or a grid, searching for a density that is closest to the ``exact" input density is usually a good idea. To do this, one could define a density error function to minimize:
	\begin{equation}\label{ErrorFunction}
		{\rm arg~min}_{v\KS(\br)}\int d\br u(\br)|n(\br)-n_{\rm in}(\br)|^w.
	\end{equation}
	subject to constraints (\ref{eqn:eigen}) and:
	\begin{equation}\label{eqn:Normal}
		\int |\psi_i(\br)|^2 d\br= 1.
	\end{equation}
	$u(\br)$ in (\ref{ErrorFunction}) can be a positive weight function that should not change the final convergence theoretically. $u(\br)=\frac{1}{n(\br)^p}$ for $p \in [1,2)$ can be helpful for the asymptotic region when $v\KS(\br)$ is calculated on a grid. Properties of the weight function $u(\br)$ when calculations are performed on a grid can be found in the recent work by Kanungo et al.\cite{kanungo2019exact}. Because in our basis-set implementation $v\KS(\br)$ is expanded as in (\ref{vKS_exp}), the asymptotic behavior is determined by $v_0(\br)$ (see below) and we use $u(\br)=1$ and $w=2$ in what follows. Note that Eq.(\ref{ErrorFunction}) is just the simplest form of an error function. The main reason to choose this form is to derive the gradient analytically, as we will do. However, with the help of automatic differentiation\cite{baydin2018automatic}, more sophisticated error functions can be designed. \par The Lagrangian can be written as
	\begin{equation}
		\label{COL}
		\begin{split}
			&L[v\KS, \{\psi_i\}, \{\epsilon_i\}, \{p_i\}, \{\mu_i\}]\\
			=& \int(n(\br)-n_{\rm in}(\br))^2 d\br \\ 
			& + \sum_{i=1}^{N/2}\int p_i(\br)(-\frac{1}{2}\nabla^2+v\KS - \epsilon_i)\psi_id\br\\
			&+\sum_{i=1}^{N/2}\mu_i(\int|\psi_i(\br)|^2d\br-1),
		\end{split}
	\end{equation}
	where $p_i(\br)$ and $\mu_i$ are the Lagrange multipliers introduced for (\ref{eqn:eigen}) and (\ref{eqn:Normal}). Variations with respect to the $\psi_i(\br)$, $\epsilon_i$ and $v\KS(\br)$ yield:
	\begin{subequations}
		\begin{align}
			\begin{split}
				&(-\frac{1}{2}\nabla^2+v\KS(\br) - \epsilon_i)p_i(\br)= 8\times \\
				&~~~(n_{\rm in}(\br)-n(\br))\psi_i(\br) - 2\mu_i\psi_i(\br),\label{CONormalEquationc}
			\end{split}\\
			&\int p_i(\br)\psi_i(\br)d\br = 0,\label{CONormalEquationd}\\
			&\frac{\delta L}{\delta v\KS(\br)}=\sum_{i=1}^{N/2}p_i(\br)\psi_i(\br).\label{CONormalEquatione}
		\end{align}
	\end{subequations}
	The optimization strategy generally adopted is to solve for the $p_i(\br)$ and the $\psi_i(\br)$ from (\ref{eqn:eigen}), (\ref{CONormalEquationc}) and (\ref{CONormalEquationd}) and build $\frac{\delta L}{\delta v_{KS}}$ from (\ref{CONormalEquatione}) for each step. The shape of $L[v_{KS}, \{\psi_i\}, \{\epsilon_i\}, \{p_i\}, \{\mu_i\}]$ can be arbitrary (not necessarily convex or concave). Thus, a good initial guess for $v\KS(\br)$ is important to find the stationary point of the Lagrangian.

	\section{Potential Basis Sets (PBS)}
	There is a general trade-off between the accuracy and efficiency in iKS. Using a PBS improves the efficiency significantly. 
	The original WY method \cite{wu2003direct} represented $v\KS(\br)$ as:
	\begin{equation}\label{vKS_exp}
		v\KS(\br) = v\ext(\br) + v_{0}(\br) + v_{\rm PBS}(\br)
	\end{equation}
	where $v\ext(\br)$ is the external potential due to the 
	nuclei and $v_{0}(\br)$ is a guide potential, whose role is discussed below. The rest is expanded on a finite potential basis set (PBS) $\{\phi_t\}$
	\begin{equation}\label{vKS_rest}
		v_{\rm PBS}(\br)=\sum_t b_t \phi_t(\br).
	\end{equation}
	Compared to a fine mesh, this choice is usually limited at resolving fine features of the XC potentials, but it greatly improves the efficiency. 
	
	The PDE-CO method has been implemented on finite-difference/finite-element meshes providing remarkably accurate results \cite{jensen2018numerical, kanungo2019exact}. Here we implement it for the first time on finite PBS and the pros and cons of finite PBS can be directly assessed.\par
	In WY, the gradient vector of $W[\Psi_{\rm det}[v\KS], v\KS]$ with respect to the $\{b_t\}$ is
	\begin{subequations}
		\label{grad}
		\begin{align}
			&\frac{\partial W[\Psi_{\rm det}[v\KS], v\KS]}{\partial b_t}\\
			= &\int d\br' \frac{\delta W[\Psi_{\rm det}[v\KS], v\KS]}{\delta v\KS(\br')}\frac{\partial v\KS(\br')}{\partial b_t} \label{grad1}\\
			=&\int d\br' (n(\br')-n_{\rm in}(\br'))\phi_t(\br')\label{grad2}.
		\end{align}
	\end{subequations}
	The Hessian matrix from first-order perturbation theory is
	\begin{subequations}
		\label{hess}
		\begin{align}
			&\frac{\partial^2W[\Psi_{\rm det}[v\KS], v\KS]}{\partial b_t\partial b_p}\\
			\begin{split}
				=&\int d\br'd\br'' \left(\frac{\delta^2 W[\Psi_{\rm det}[v\KS], v\KS]}{\delta v\KS(\br')v\KS(\br'')}\times\right.\\
				&\left.\frac{\partial v\KS(\br')}{\partial b_t}\frac{\partial v\KS(\br'')}{\partial b_p}\right)   \label{hess1}
			\end{split}
			\\=& 2\sum_i^{occ.}\sum_a^{unocc.}
			\frac{<\psi_i|\phi_t|\psi_a><\psi_a|\phi_p|\psi_i>}{\epsilon_i - \epsilon_a} \label{hess2}.
		\end{align}
	\end{subequations}
	It can be easily proven that the Hessian matrices are negative definite, which means that $W[\Psi_{\rm det}[v\KS], v\KS]$ is concave for any given PBS $\{\phi_t\}$ as discussed above.\par
	Table \ref{GradientCheck} shows that both gradient and Hessian, calculated through a finite-difference approximation, are numerically accurate. The tests reported on Table \ref{GradientCheck} also show where the influence of numerical errors comes in. It can be seen that, before the optimization, the relative errors for the gradient are always small. However, after convergence has been achieved and the analytical gradients are small, the relative gradient errors increase because the absolute numerical errors remain.\par
	\begin{table*}[t]
		\centering
		\caption{Wu-Yang: Finite Difference Tests}
		\begin{tabular}{|c||c|c|c|c|c|c|}
			\hline
			\multicolumn{7}{|c|}{Wu-Yang Gradient Test\textsuperscript{\emph{a}}} \\
			\hline
			Basis Set & \multicolumn{2}{|c|}{CD/CD}  & \multicolumn{2}{|c|}{CD/CQ} & \multicolumn{2}{|c|}{CQ/CQ}\\
			\hline
			&before\textsuperscript{\emph{c}} &after &before &after &before &after\\
			\hline
			Be &5.6e-6  &4.1e-3  &3.9e-6  &0.5  &1.8e-5  &1.3e-2 \\
			\hline
			Ne &5.5e-6 &0.06 &5.8e-6 &2.0e-2 &5.3e-5 &2.4e-2\\
			\hline
			Ar &1.6e-5 &0.2 &3.9e-5 &0.24 &5.9e-5 &0.2\\
			\hline
		\end{tabular}
		\begin{tabular}{|c||c|c|c|c|c|c|}
			\hline
			\multicolumn{7}{|c|}{Wu-Yang Hessian Test\textsuperscript{\emph{b}}} \\
			\hline
			Basis Set & \multicolumn{2}{|c|}{CD/CD}  & \multicolumn{2}{|c|}{CD/CQ} & \multicolumn{2}{|c|}{CQ/CQ}\\
			\hline
			&before &after &before &after &before &after\\
			\hline
			Be &2.3e-7 &8.1e-8 &1.5e-7 &7.0e-8 &1.5e-6 &1.8e-7\\
			\hline
			Ne &2.2e-7 &3.2e-8 &3.1e-7 &6.3e-8 &2.6e-6 &1.5e-7\\
			\hline
			Ar &4.3e-7 &8.8e-8 &6.6e-7 &1.0e-7 &2.7e-6 &3.2e-7\\
			\hline
		\end{tabular}
		\\\textsuperscript{\emph{a}} Relative errors for gradient ($\|grad-grad_{approximation}\|/\|grad\|$);
		\\\textsuperscript{\emph{b}} Relative errors for Hessian ($\|hess-hess_{approximation}\|/\|hess\|$);
		\\\textsuperscript{\emph{c}} ``before" means result before the optimization; ``after" means result after the optimization.
		
		\label{GradientCheck}
	\end{table*}
	
	For the PDE-CO method, we represent all the terms on finite basis sets. The KS-orbitals $\psi_i(\br)$ are represented on a finite OBS and $v\KS(\br)$ is represented on PBS as discussed above. We use $\{\phi\}$ to denote the PBS and $\{\phi'\}$ for the OBS. The $p_i(\br)$ are expanded on the OBS, just like $\psi_i(\br)$:
	\begin{equation}
		\label{piExpension}
		p_i(\br) = \sum_k c_{ik}\phi_k'(\br).
	\end{equation}
	To solve for the coefficients $c_{ik}$, we multiply $\phi_k'$ on both sides of (\ref{CONormalEquationc}) and integrate:
	\begin{equation}
		\label{solve_p}
		(\mF-\epsilon_i\mS)\vc_i = \vg_i
	\end{equation}
	where $\mF$ is the Fock matrix and $\mS$ is the overlap matrix $\mS_{ij}=\langle\phi_i'|\phi_j'\rangle$.
	The vectors $\vg_i$,
	\begin{equation}
		\label{Gmatrix}
		\{\vg_i\}_k = \int d\br \phi_k'(\br)[8(n_{\rm in}(\br)-n(\br))\psi_i(\br) - 2\mu_i\psi_i(\br)],
	\end{equation}
	are derived from (\ref{CONormalEquationc}).  The gradient is derived from (\ref{CONormalEquatione}):
	\begin{equation}
		\label{PDEGradient}
		\begin{split}
			\frac{dL}{db_i} 
			&= \int d\br \frac{\delta L}{\delta v\KS(\br)} \phi_i(\br) \\
			&= \sum_{j=1}^{N/2} \int d\br p_j(\br)\psi_j(\br)\phi_i(\br).
		\end{split}
	\end{equation}
	
	The well-known limitations of finite basis sets in connection to the ill-posed nature of iKS have stimulated the development of
	many methods \cite{bulat2007optimized, gaiduk2013removal, heaton2008optimized, jacob2011unambiguous, wu2003direct, nam2020kohn} to improve the results of density-to-potential inversions, though some of them have overlapped effects and they all have shortcomings. To date, there is no clear and straightforward strategy that works reliably. In the following, we explain the theoretical effects of basis sets, regularization, guide potentials and use of different optimizers, compare the results, and provide guidance on how to utilize them. In the end, considering both accuracy and efficiency, we provide a recommendation that is most robust (though not perfect) according to our experience, and could be directly adopted for practical calculations.\par
	
	\section{Orbital Basis Sets (OBS)}
	The limitations of finite OBS in density-to-potential inversion are well documented \cite{mura1997accurate, schipper1997kohn, staroverov2006optimized}. 
	Moreover, since atomic orbitals are specifically designed to describe molecular orbitals, there is no guarantee that they can be satisfying for PBS, which is often true for inversion in embedding methods, whose embedding potentials usually have different spatial features compared to molecular orbitals. 
	In practice, the results of both WY and PDE-CO methods can be highly sensitive to both the OBS and the PBS, but 
	there is usually very little freedom in choosing them. 
	There are two main reasons finite OBS can lead to unphysical oscillations in the potential. First, for a given finite OBS $\{\phi_{\mu}'\}$, all the XC potentials that produce the same Fock matrices:
	\begin{equation}\label{vxcFock}
		V_{\mu\nu} = \int \phi_{\mu}'^*(\br)v(\br)\phi_{\nu}'(\br)d\br,
	\end{equation}
	will lead to the same density. This many-to-one problem is at the root of the relation between the exact external potential and `effective' external potentials discussed by Gaiduk et al.\cite{gaiduk2013removal} and Kumar et al.\cite{kumar2020accurate}. Second, the electron density cannot be represented exactly on finite Gaussian basis sets. Small input errors will lead to large oscillations given the ill-posed nature of iKS.\cite{schipper1997kohn, jensen2018numerical}\par
\begin{figure}[h!]
		\centering
		\includegraphics[width=3.3in]{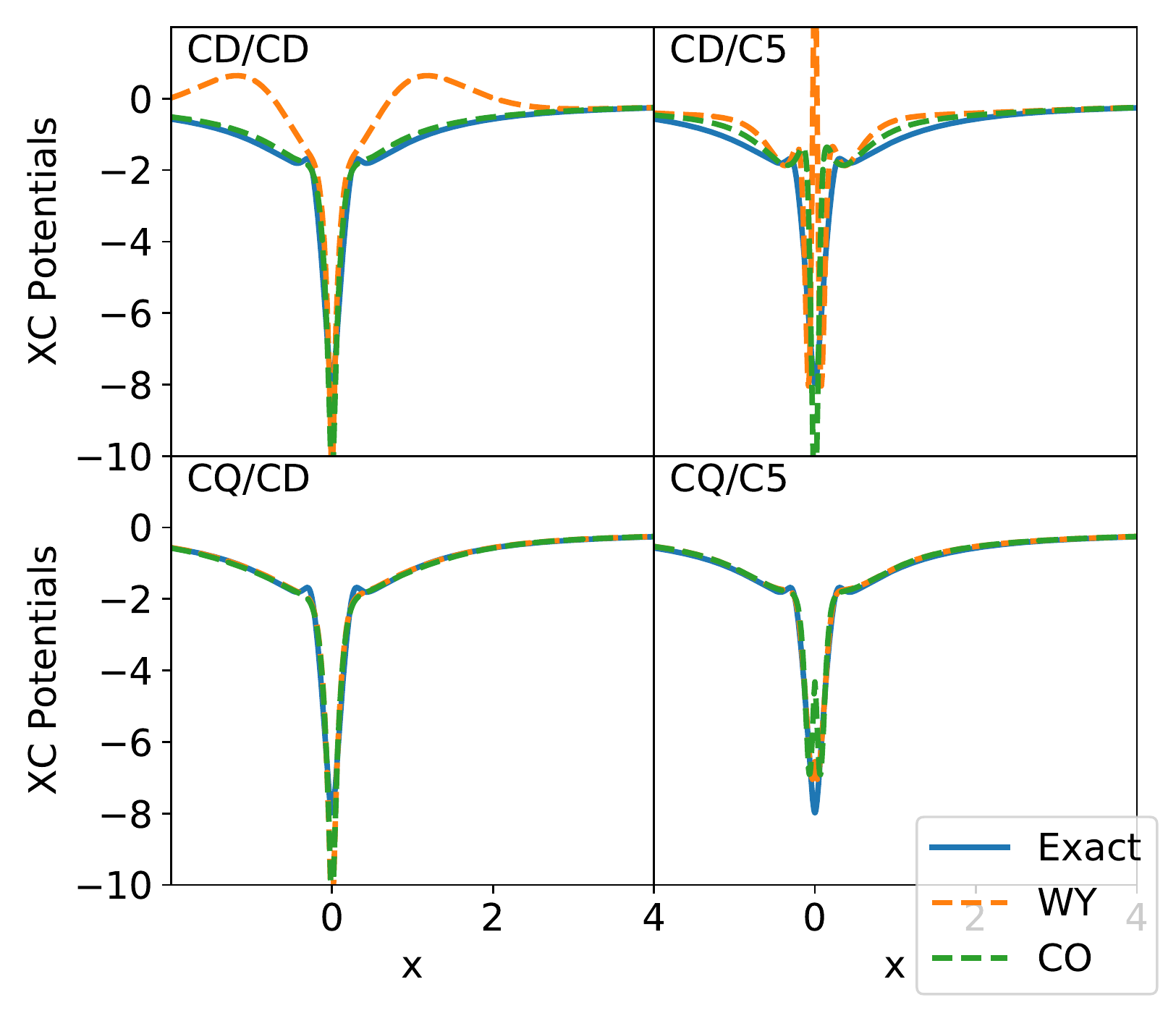}
		\caption{Neon atom XC potential from different combinations of OBS and PBS. BFGS runs for 30 iterations for all the results.}
		\label{fig:BasisSets}
	\end{figure}

	To address the first point above, we find that using similar basis sets, even small ones, for orbital and potential is typically a good choice. There is often error around the nuclei (Fig. \ref{fig:BasisSets}), where both OBS and PBS are insufficiently accurate. 
	On the other hand, even though large basis sets for both the orbital and potential are preferred, this is impractical. A well-adopted strategy is to use a relatively small OBS to reduce the computational cost and to use a large and carefully chosen PBS to have a good representation for the potential. Following this strategy often introduces unphysical oscillations and thus regularization methods are required.\par
	
\begin{figure}[h!]
		\centering
		\includegraphics[width=1\linewidth]{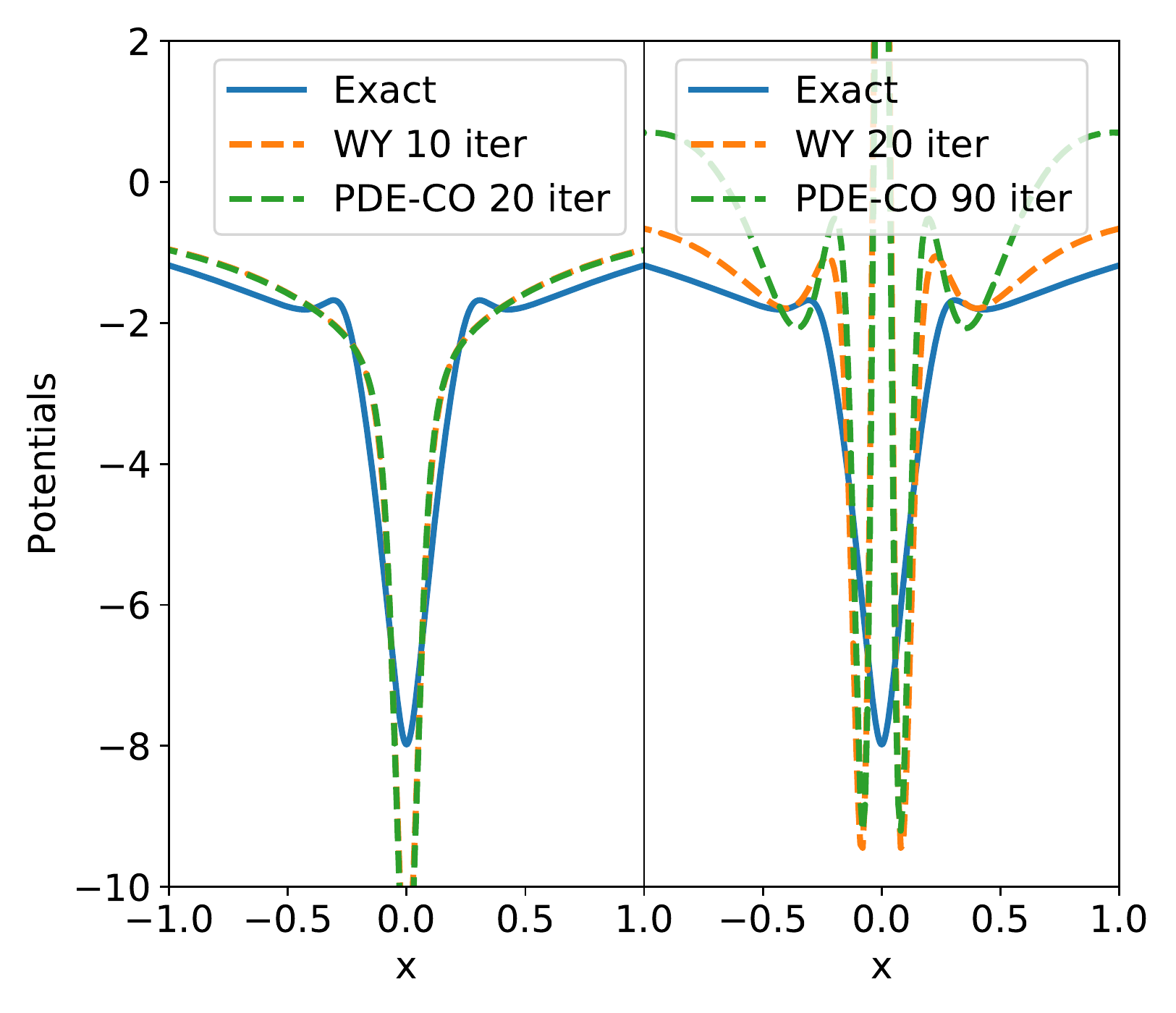}
		\caption{Ne atom XC potentials on CD/C5. BFGS is used with different number of iterations specified. The same initial point ($\{b_t\}=0$ and $v_0 = v_{\rm FA}$) is used.}
		\label{fig:BasisSet2}
	\end{figure}

	To better illustrate the limitations of finite basis sets and the sensitivity of the resulting potentials, we compare in Fig. \ref{fig:BasisSets} the XC potentials obtained for the Ne atom from WY and PDE-CO methods using different combinations of basis sets. 
	The results are very sensitive to the choice of PBS and OBS.
	One could imagine that the unphysical features of the resulting XC potentials would depend strongly on the method used to perform the iKS. However, Fig.\ref{fig:BasisSet2} confirms that they are not. Even though the WY and PDE-CO methods are based on different principles, with optimizers traveling along different paths in the space spanned by the $\{\phi_t\}$, the unphysical features of the resulting XC potentials are largely the same, confirming the dominant role played by the finite basis sets.
	
	\section{Guide Potentials} 
\begin{figure}[h!]
		\centering
		\includegraphics[width=1\linewidth]{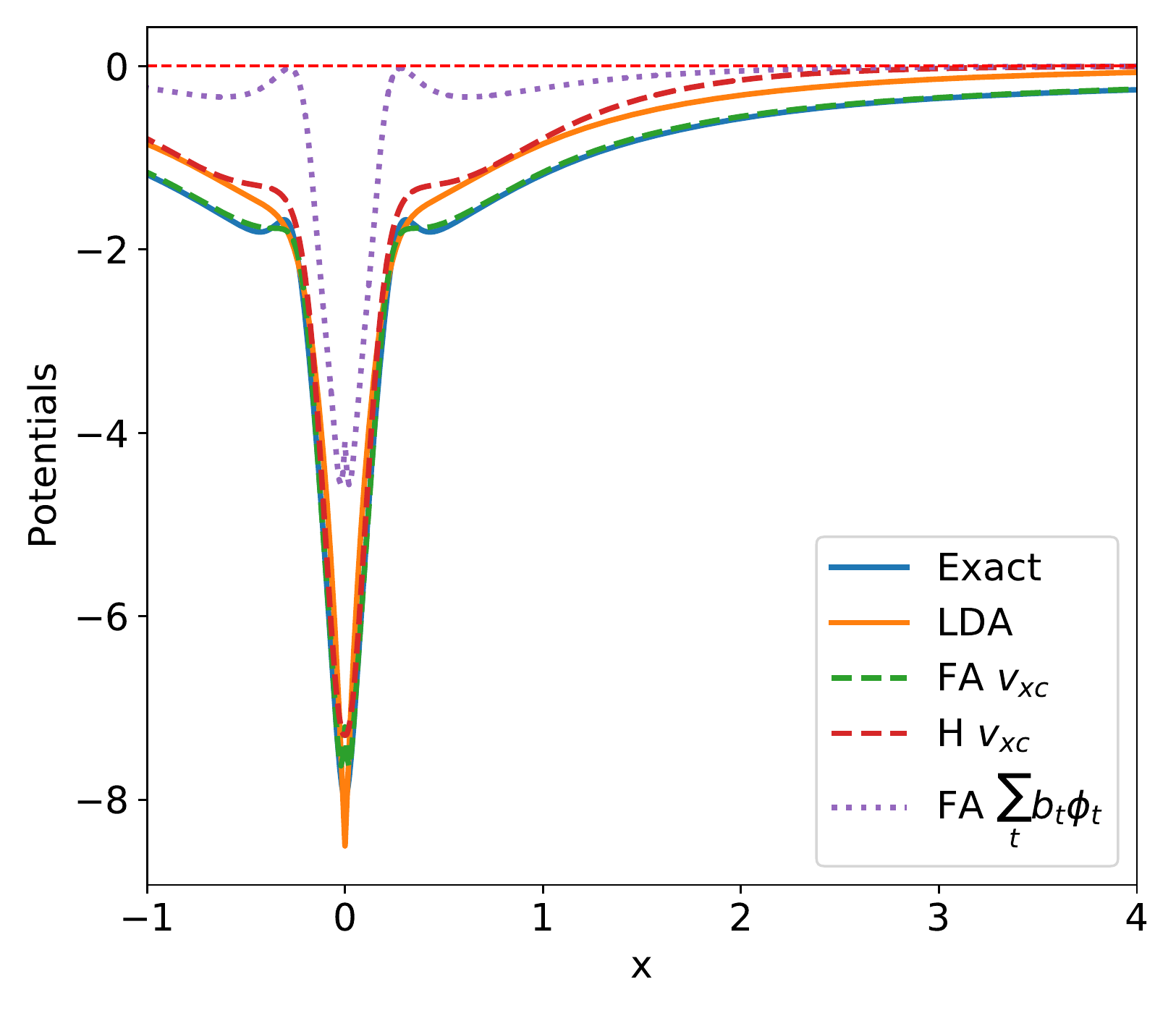}
		\caption{Neon atom XC potential and $\sum_t b_t \phi_t(\br)$ for different $v_0$ calculated in C5/C5. FA is Fermi-Amaldi and H is Hartree. The red dashed line is $y=0$.}
		\label{v0_Fig}
	\end{figure}

	The WY original method \cite{wu2003direct} uses the Fermi-Amaldi (FA) potentials \cite{wu2003direct, zhao1994electron}
	\begin{equation}\label{FA}
		v_{\rm FA}(\br) = \frac{N-1}{N} \int d\br' \frac{n_{\rm in}(\br')}{|\br-\br'|}
	\end{equation}
	as a guide potential $v_0(\br)$ in Eq.(\ref{vKS_exp}), following \citeauthor{zhao1994electron} \cite{zhao1994electron}. 
	Since the FA potential can be understood as a Hartree potential $
	v\H(\br) = \int d\br' \frac{n_{\rm in}(\br')}{|\br-\br'|}$ with a part from exchange-correlation that partially prevents self-interaction, we also test $v_0(\br)=v\H(\br)$ below. 
	Because PBS usually has a poor behavior in the asymptotic region, $\sum_t b_t \phi_t(\br)$ often decays to $0$ too quickly (see Figure \ref{v0_Fig}). In other words, $v_0(\br)$ should have the asymptotic behavior of $v\H(\br)+v\xc(\br)$ because $\sum_t b_t \phi_t(\br)$ has negligible contributions far from the nuclei. In the case of LDA densities, $v_0(\br) = v\H(\br)$ fits this requirement better. However, the XC potentials obtained using $v_0(\br) = v_{\rm FA}(\br)$ usually have an asymptotic behavior that runs much closer to the exact one and leads to better satisfaction of Koopmans' theorem, $\epsilon_{\rm HOMO}=-I$. This is true even when the LDA input density is used \cite{wu2003density, callow2020density}. 
	
	\section{Regularization} 
\begin{figure}[h!]
		\centering
		\includegraphics[width=3.3in]{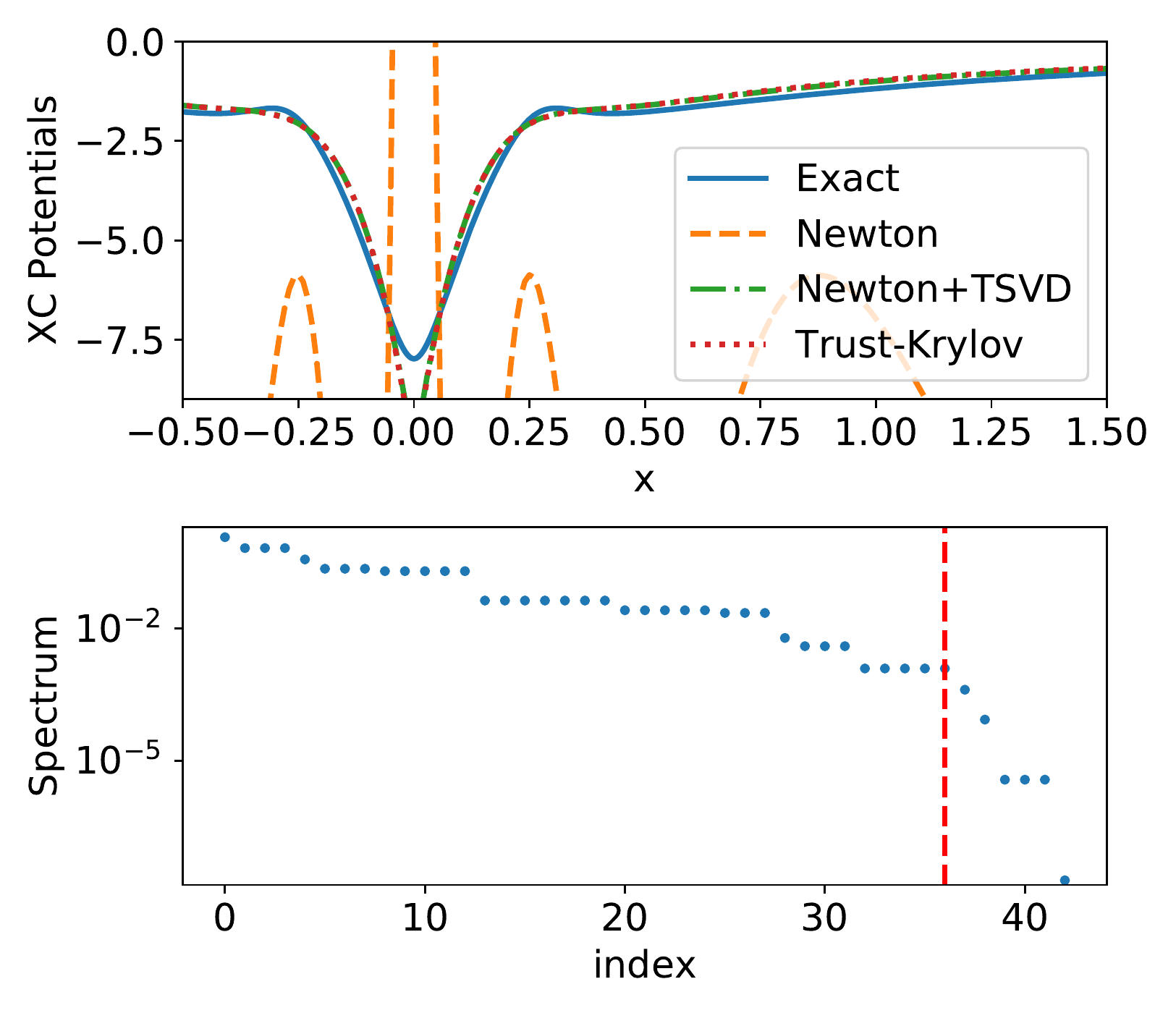}
		\caption{Neon atom XC potential using Newton+TSVD. The top panel shows results from different optimization methods. The bottom graph is the spectrum of singular values of the Hessian matrix before the convergence. The vertical red dashed line is the TSVD cutoff. CD/CT basis sets are used.}
		\label{TSVD}
	\end{figure}

	The effects of regularization are usually significant. The response function in (\ref{equ:WYHess}) and the Hessian matrix in (\ref{hess2}) are supposed to sum over all unoccupied KS orbitals. The real response functions are invertible, but the Hessian matrices approximated with a finite number of KS orbitals are usually very singular, especially when the PBS differs significantly or is larger than the OBS \cite{gidopoulos2012nonanalyticity}, as shown in Fig. \ref{TSVD}. truncated singular value decomposition (TSVD) regularization is thus often necessary. \citeauthor{wu2003direct} used TSVD \cite{hansen1987truncatedsvd} 
	without providing a systematic way to determine the truncation threshold \cite{wu2003direct}. \citeauthor{bulat2007optimized} introduced an additional $\lambda$-regularizer
	\begin{equation}\label{lambda_regularizer}
		-\lambda \|{\nabla v_{\rm PBS}(\br)}\|^2 \rightarrow -2\lambda\vb^T\mT\vb
	\end{equation}
	where $\vb$ is the coefficient vector for $v_{\rm PBS}(\br)$ and $\mT$ is the kinetic contribution to the Fock matrix \cite{bulat2007optimized}. 
	The L-curve method was introduced to search for the regularization parameter $\lambda$ \cite{bulat2007optimized, heaton2008optimized}. Both methods are widely used as standard regularization for numerical optimization. They both contribute to the stability of the inversions irrespective of the basis-set employed and help overcome the over-fitting problems especially around the nuclei. To demonstrate this, we introduce here two simple but effective tricks: {\bf (1)} Cliff-plotting to determine the optimum parameters for TSVD; and {\bf (2)} $T_s$-focusing to optimize $\lambda$ for $\lambda$-regularization.\par
	
	{\bf (1)} The truncating parameters for TSVD depend on the systems and basis sets. The most straightforward method to determine these parameters is to plot the spectrum (the diagonal vector of the S matrix in SVD, see Figure \ref{TSVD}). In many cases, the spectrum is of a ``cliff" shape. Cutting at the edge of the cliff usually works. The main idea is to eliminate the subspace whose variations do not change the Fock matrices in Eq.(\ref{vxcFock}). This simple method can be tricky when the OBS is much smaller than the PBS, where the spectrum shows several cliffs or disconnections and one has to decide on which one to cut. On the other hand, when the OBS and PBS are very close or the same, TSVD is often unnecessary. Moreover, optimizers like Trust-Krylov \cite{gould1999solving}
	and L-BFGS \cite{byrd1995limited}
	yield similar result as the Newton method with TSVD (Figure \ref{TSVD}).\par
\begin{figure}[h!]
		\centering
		\includegraphics[width=3.3in]{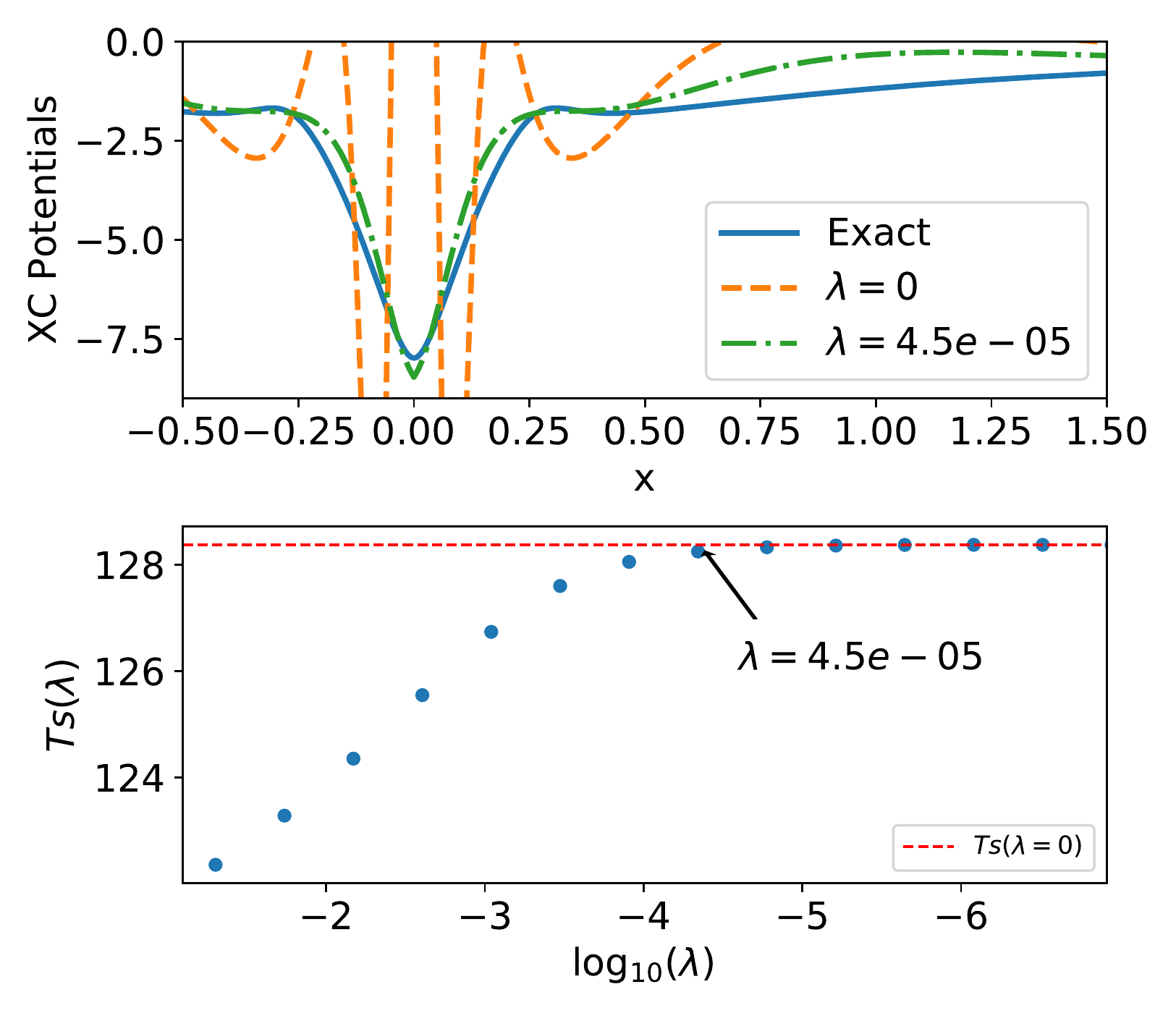}
		\caption{Neon atom XC potentials from the Wu-Yang method with $\lambda$-regularization. Newton with TSVD is used on CD/C5.}
		\label{LCurveFig}
	\end{figure}

\begin{figure}[h!]
		\centering
		\includegraphics[width=1\linewidth]{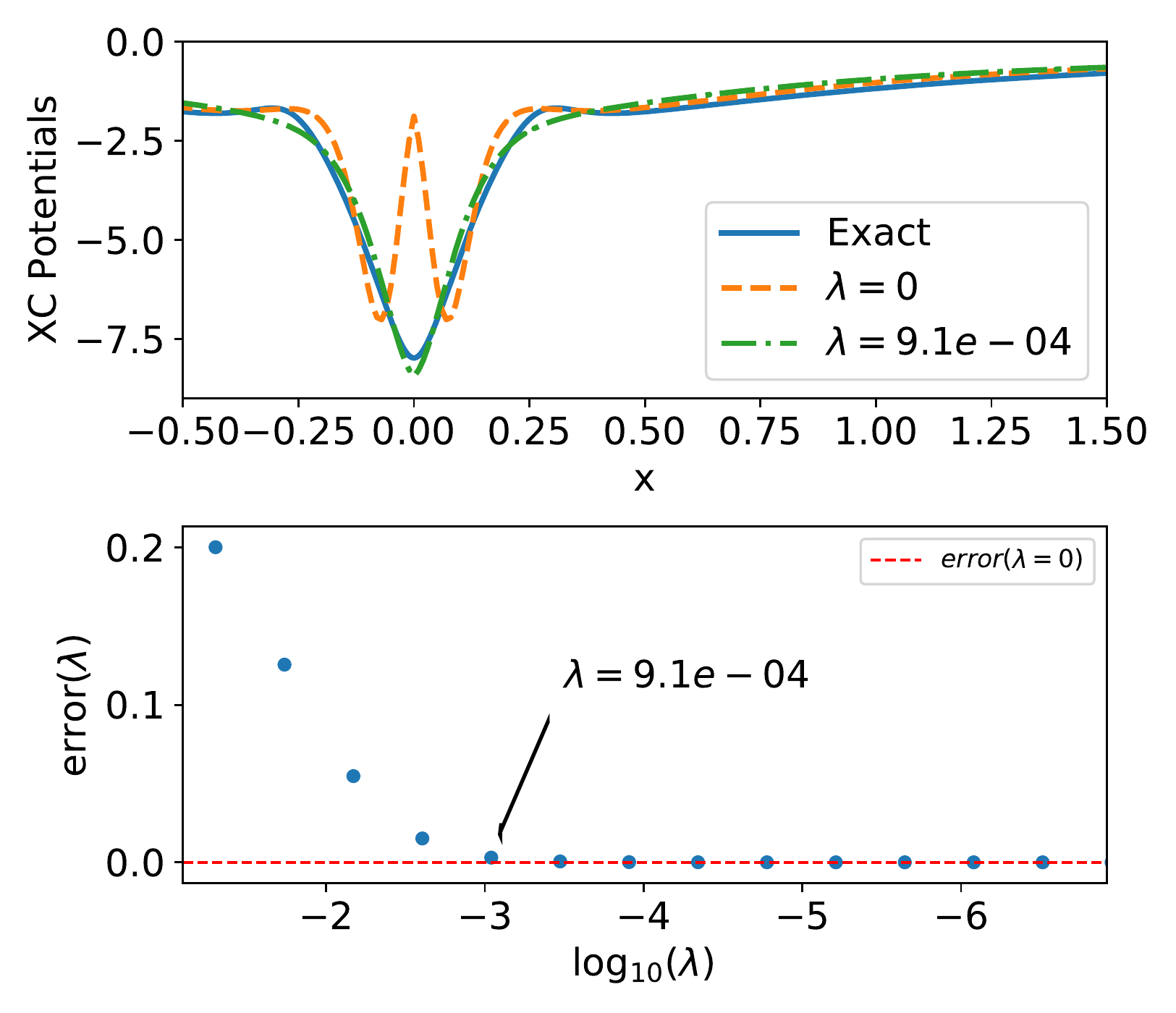}
		\caption{Neon atom XC potentials from the PDE-Constrained optimization with $\lambda$-regularization. $error(\lambda)$ is the density error function defined in (\ref{ErrorFunction}). L-BFGS is used on CD/C5.}
		\label{CO_lambda_regularization}
	\end{figure}

	{\bf (2)} Even with TSVD, there can still be over-fitting, especially around the nuclei (see Figure \ref{TSVD}). We find that $\lambda$-regularization is more reliable in this region.
	In addition to the L-curve method introduced by \citeauthor{bulat2007optimized} \cite{bulat2007optimized}, we test here an alternative method to search for $\lambda$. We calculate $T_s(\lambda)$ as a function of $\lambda$ (see Figure \ref{LCurveFig}). The point of $T_s(\lambda)$ with the largest $\lambda$ that is close enough to $T_s(\lambda=0)$ is chosen. The main idea is to adopt the simplest potential (corresponding to the largest $\lambda$) which does not essentially change the optimized result.
	Though obtaining $T_s(\lambda)$ curves such as those of Figs. \ref{LCurveFig}-\ref{CO_lambda_regularization} requires multiple inversion calculations, the efficiency of constrained optimization methods on finite PBS makes it acceptable in practice.\par
	Motivated by the very similar overfitting features exhibited by the WY and PDE-CO methods, as evidenced in Fig.\ref{fig:BasisSet2}, we now 
	add $\lambda$-regularization to the Lagrangian of the PDE-CO method, Eq.(\ref{COL}). Note the sign for the regularization term needs to be changed when compared to (\ref{lambda_regularizer}) because (\ref{COL}) is minimized while (\ref{equ:WuYangL}) is concave for a given PBS. We follow a similar strategy to search for $\lambda$ as we have done for WY. Rather than $T_s(\lambda)$, the error (\ref{ErrorFunction}) as a function of $\lambda$ is utilized and the largest $\lambda$ for a small enough error is selected (Fig. \ref{CO_lambda_regularization}). In general it is impractical to search for a perfect $\lambda$ for a specific OBS/PBS. The simple method introduced above usually effectively eliminates the oscillations.
	
	\section{Null-Space Correction (NSC):} 
	When using TSVD, it is possible that there is not a single clear edge and cutting at one edge does not lead to a smooth result. Also, cutting a Hessian matrix in this way ``wastes" the large PBS selected. Instead of $\lambda$-regularization, a correction to the Hessian matrix can be defined based on the Uns\"old approximation \cite{unsold1927quantentheorie}. 
	This follows the derivation given by \citeauthor{gidopoulos2012nonanalyticity} \cite{gidopoulos2012nonanalyticity} for OEP.
	When the Hessian information is used for the Wu-Yang method optimization, the following equation:
	\begin{equation}\label{Newtonsequation}
		\mH_0 \vb_0 = \boldsymbol{g}
	\end{equation}
	is solved, where $\vb_0$ is composed of the update coefficients for $v_{\rm PBS}(\br)$: 
	\begin{equation}\label{potential_update}
		\begin{split}
			v^{k+1}_{\rm PBS}(\br)
			&= v^{k}_{\rm PBS}(\br) + \bigtriangleup v^{k}_{\rm 0,PBS}\\
			&= v^{k}_{\rm PBS}(\br) + \sum_t b_{0,t}\phi_t(\br),
		\end{split}
	\end{equation}
	and $\boldsymbol{g}$ is the gradient vector. The null-space vector \(\bar{\vb}\) of the Hessian matrix \(\mH_0\) is added:
	\begin{equation}\label{NewtonsequationNullSpace}
		\mH_0 (\vb_0 + \bar{\vb}) = \boldsymbol{g},
	\end{equation}
	i.e. $\mH_0 \bar{\vb} = 0$. Now we define a $\bar{\vb}$ that is a correction to $\vb_0$ similar to the derivation by \citeauthor{gidopoulos2012nonanalyticity}  \cite{gidopoulos2012nonanalyticity}, where the correction is derived for the Optimized Effective Potential. \par
	The real response function $\chi(\br, \br')$ is invertible; we denote the rest as $\Tilde{\chi}(\br,\br')$,
	\begin{equation}\label{ResponseTilde}
		\Tilde{\chi}(\br,\br') = \chi(\br,\br') - \chi_0(\br,\br')
	\end{equation}
	where $\chi_0$ is again the approximation summed over all the virtual orbitals in the KS-Slater determinant corresponding to \(\mH_0\) on finite basis sets. Using a short-hand notation in which \(\chi\bigtriangleup\vv\) stands for \(\int d\br' \chi(\br, \br') \bigtriangleup\vv(\br')\), we have
	\begin{equation}\label{GLlimit1}
		(\chi_0 + \Tilde{\chi})\bigtriangleup v = n-n_{\rm in}
	\end{equation}
	and further
	\begin{equation}\label{RowSpaceNewton}
		\chi_0\bigtriangleup v_0 = n-n_{in},
	\end{equation}
	The null-space vector $(\Tilde{g}(\br))$ can be separated from the right hand side, .i.e. $n(\br)-n_{in}(\br) = g(\br) + \Tilde{g}(\br)$, where $\int d\br' \chi_0(\br, \br') \Tilde{g}(\br') = 0$. A parameter $\lambda$ is added to control the null-space component:
	\begin{equation}\label{GLlimit2}
		(\chi_0 + \lambda\Tilde{\chi})\bigtriangleup v^{\lambda} = g + \lambda\Tilde{g}.
	\end{equation}
	When $\lambda=1$, this is exactly (\ref{GLlimit1}). By setting $\lambda \rightarrow 0$, $\bigtriangleup v^{\lambda \rightarrow 0}$ can be derived. The null-space correction to $v_0(\br)$ is defined as difference between $\bigtriangleup v^{\lambda=0}(\br)$ and $\bigtriangleup v^{\lambda \rightarrow 0}(\br)$
	\begin{equation}\label{vbar}
		\bigtriangleup \bar{v}(\br) = \bigtriangleup v^{\lambda \rightarrow 0}(\br) - \bigtriangleup v_0(\br).
	\end{equation}
	The discontinuity is shown as following. By setting $\lambda \rightarrow 0$, $\bigtriangleup v^{\lambda \rightarrow 0}(\br)$ can be expanded by orders of $\lambda$ using a Taylor series of $\bigtriangleup v^{\lambda}(\br)$ around $\lambda=0$, i.e. $\bigtriangleup v^{\lambda}(\br) = \bigtriangleup v^{\lambda \rightarrow 0}(\br) + \lambda \bigtriangleup v' + \dots$. By matching the order of $\lambda$:
	\begin{subequations}
		\label{GLlimitExpension}
		\begin{align}
			\mathcal{O}(1): &\chi_0 \bigtriangleup v^{\lambda \rightarrow 0} = g \label{GLlimitExpensionConstant}\\
			\mathcal{O}(\lambda): &\chi_0 \bigtriangleup v' + \Tilde{\chi} (\bigtriangleup \bar{v} + \bigtriangleup v_0) = \Tilde{g}.\label{GLlimitExpensionLinear}
		\end{align}
	\end{subequations}
	(\ref{GLlimitExpensionConstant}) and (\ref{RowSpaceNewton}) together prove that $\bigtriangleup \bar{v}$ is the null space vector of $\chi_0$.\par
\begin{figure}[h!]
		\centering
		\includegraphics[width=3.3in]{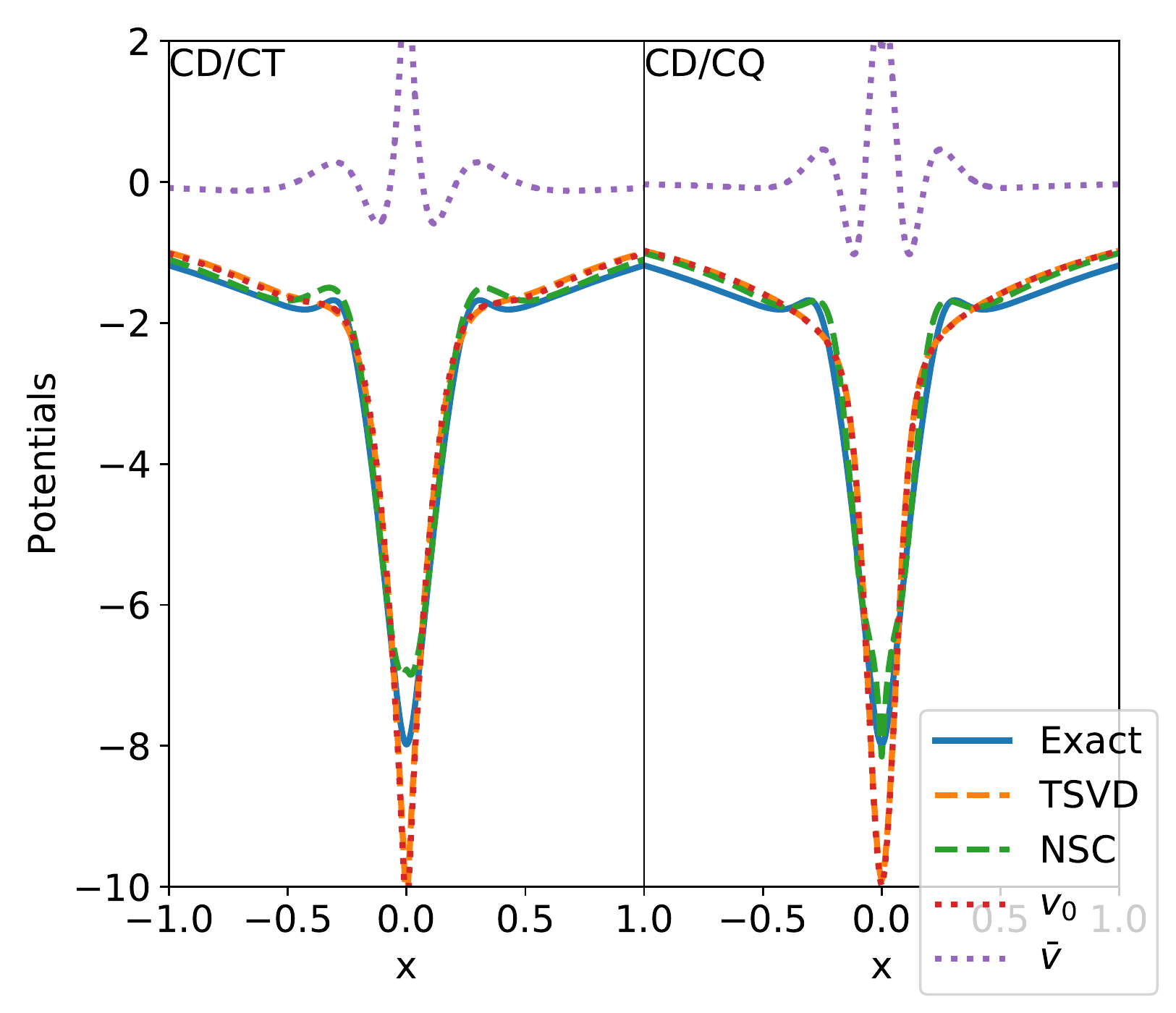} 
		\caption{Neon atom potentials in CD/CT (left) and CD/CQ (right). TSVD and Null-Space Corrected (NSC) XC potentials are compared. $v_0$ is the original result and $\bar{v}$ is the correction term (i.e. ${\rm NSC}=v_0+\bar{v}$).}
		\label{GL12}
	\end{figure}

	By expanding all the potentials on the finite potential basis sets and integrating with the null-space vectors, the first term in (\ref{GLlimitExpensionLinear}) disappears and it turns into matrix form:
	\begin{equation}\label{ChiTilde}
		(\mC^{\mu T} \tilde{\mH}) (\vb_0 + \bar{\vb}) = \mC^{\mu T}\boldsymbol{\tilde{g}}.
	\end{equation}
	and
	\begin{subequations}
		\label{ResponseEigenNull}
		\begin{align}
			\mH_0 \vc^{\alpha} &= \epsilon_{\alpha}\vc^{\alpha}\\
			\mH_0 \vc^{\mu} &= 0.
		\end{align}
	\end{subequations}
	In order to do this, TSVD is used. We adopt the notation $\mU^{\alpha}, \mV^{\alpha}$ and $\mU^{\mu}, \mV^{\mu}$ for both $\epsilon^{\alpha} \leq \epsilon^{cutoff}$ and $\epsilon^{\mu} > \epsilon^{cutoff}$.
	$\boldsymbol{\tilde{g}}$ is $n(\br) - n_{\rm in}(\br)$ projection in the null space and integrated with the PBS:
	\begin{equation}\label{gradTilde}
		\boldsymbol{\tilde{g}} = \mU^{\mu} \mV^{\mu\mT} \boldsymbol{g}
	\end{equation}
	To approximate for $\tilde{\mH}$, the Uns\"old approximation \cite{unsold1927quantentheorie} is adopted by setting the energy difference as a constant $C \simeq \epsilon_a - \epsilon_i$:
	\begin{equation}\label{ChiTilde}
		\begin{split}
			C \times \tilde{H}_{mn}
			=&\sum_i <\psi_i|\phi_m\phi_n|\psi_i> \\&-
			\sum_i \sum_{p=\{i,a_0\}}
			<\psi_i|\phi_m|\psi_p><\psi_p|\phi_n|\phi_i>
		\end{split}
	\end{equation}
	where $a_0$ denote the finite set of KS unoccupied orbitals.
	Since $\bar{v}(\br)$ is a null space function of $\chi_0$, it can be written as a linear combination of the null-space basis $\{u^{\nu}(\br)\}$:
	\begin{equation}\label{d-bar}
		\bar{\vb} = \mV^{\mu}\bar{\vp}
	\end{equation}
	and one gets
	\begin{equation}\label{p-bar}
		\mU^{\mu T}\tilde{\mH}\mV^{\mu}\bar{\vp} = \mU^{\mu T} \boldsymbol{\tilde{g}} -
		\mU^{\mu T} \tilde{\mH}\vb_0
	\end{equation}
	then
	\begin{equation}\label{p-bar1}
		\bar{\vp} = \mA^{-1}\mU^{\mu T} \boldsymbol{\tilde{g}} -
		\mA^{-1}\mU^{\mu T} \tilde{\mH}\vb_0
	\end{equation}
	where $\mA = \mU^{\mu T}\tilde{\mH}\mV^{\mu}$. This is similar to Eq.(50) given by \citeauthor{gidopoulos2012nonanalyticity} \cite{gidopoulos2012nonanalyticity}. Because $\chi$ is supposed to be in general invertible, $\tilde{\chi}$'s projection into $\chi_0$'s null space, $\mA$, should be invertible if the exact $\tilde{\mH}$ is used. This is not usually true in practice. TSVD is often necessary for the inversion of $\mA$. This correction fixes unphysical oscillations well, as shown in Fig. \ref{GL12}.
	This correction does not contribute in some cases, especially when the OBS and PBS are similar.
	\section{Optimization Methods} 
	The choice of optimization methods can also be important. Even for the WY method, which is concave, the optimized results can be very different with similar $T_s$ and density error (\ref{DensityError}) as we mentioned above. Here, we compare the performance of some of the most widely used methods that are available from the Scipy.optimize library \cite{virtanen2020scipy}: quasi-Newton (BFGS \cite{fletcher2013practical}, L-BFGS-B \cite{byrd1995limited}), and trust-region methods (exact \cite{conn2000trust}, Krylov \cite{gould1999solving}). 
	We also implement Newton with TSVD regularization manually. As Wu and Yang pointed out\cite{wu2003direct}, the BFGS does not require TSVD and can often give a better convergence regarding $W[\Psi_{\rm det}[v\KS], v\KS]$ though more iterations are required.\par
	We consider four different criteria including $W[\Psi_{\rm det}[v_{\rm KS}], v_{\rm KS}]$, $T_s[\Psi_{\rm det}]$, the gradient (\ref{grad}) norm $\|grad\|$ and density error
	\begin{equation}\label{DensityError}
		N_{error} = \int d\br|n(\br) - n_{\rm in}(\br)|.
	\end{equation}
	The gradient norm $\|grad\|$ can be understood as metrics of optimization on a given PBS. The density error is more general and unlimited by the PBS. The gradient norm optimization is a necessary but insufficient condition for density error optimization because of the limitation of finite PBS (Table \ref{twoErrors}). The gradient norm is the main criterion used by optimizers for convergence.\par
	\begin{table*}[t]
		\centering
		\caption{Two Different Norm Comparison at convergence.}
		\begin{tabular}{|c||c|c|c|c|c|c|}
			\hline
			Basis Set & \multicolumn{2}{|c|}{CD/STO-3G}  & \multicolumn{2}{|c|}{CD/CD} & \multicolumn{2}{|c|}{CD/CQ}\\
			\hline
			&$\|grad\|$ &$N_{error}$ &$\|grad\|$ &$N_{error}$ &$\|grad\|$ &$N_{error}$\\
			\hline
			Be &7.5e-6 &1.8e-2 &4.8e-5 &1.6e-2 &8.8e-7 &4.2e-6 \\
			\hline
			Ne &2.7e-6 &3.1e-2 &1.5e-5 &7.8e-4 &5.9e-5 &1.5e-3 \\
			\hline
			Ar &5.5e-8 &9.2e-2 &2.1e-6 &6.8e-2 &5.2e-5 &4.7e-3\\
			\hline
		\end{tabular}
		\label{twoErrors}
	\end{table*}
	
\begin{figure}[h!]
		\centering
		\includegraphics[width=3.3in]{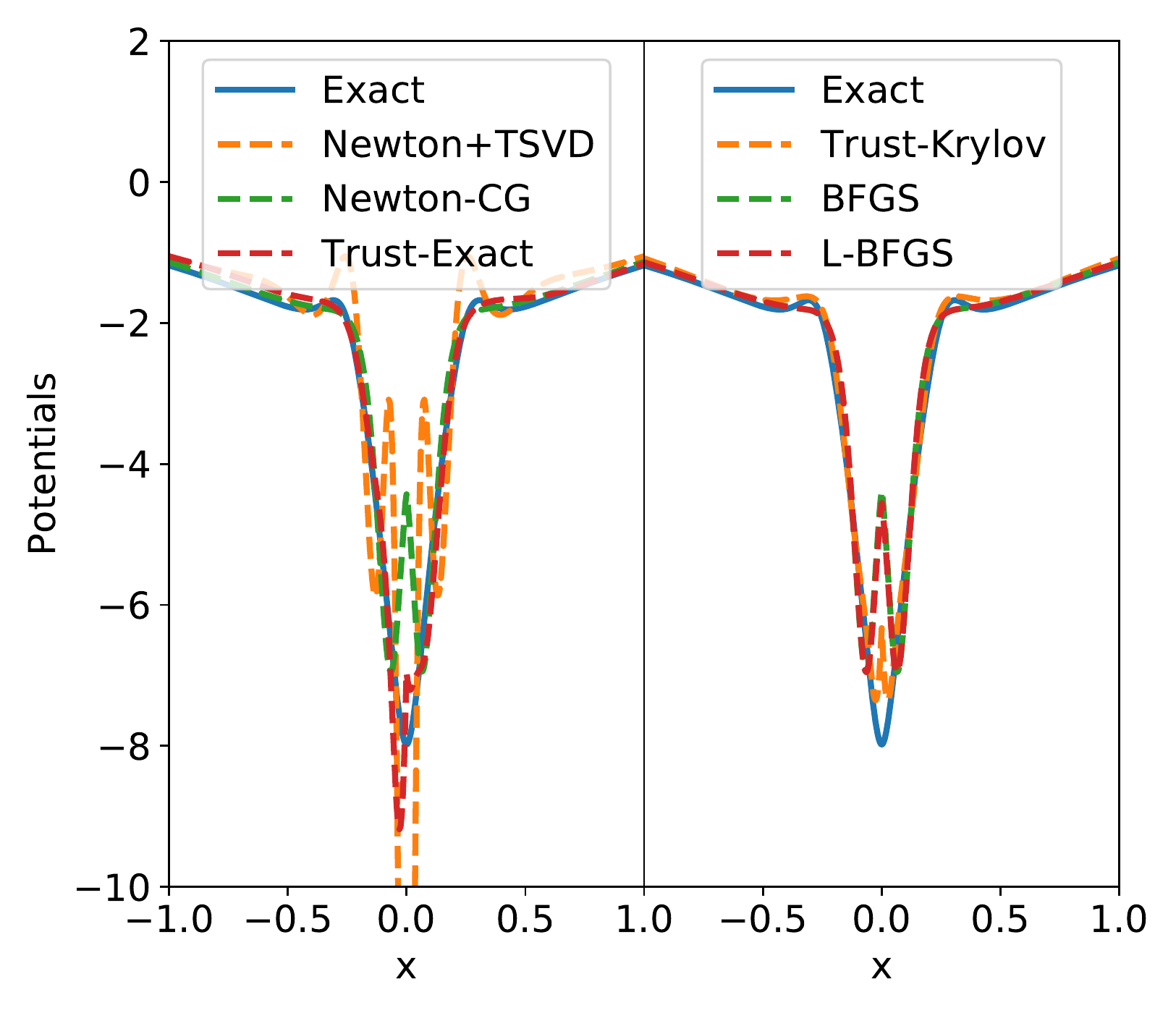}
		\caption{Neon atom XC potentials for common optimization methods in CT/C5.}
		\label{OptMedFig}
	\end{figure}

	Table \ref{OptMedTable} and Figure \ref{OptMedFig} illustrate the performance of these different methods on the Ne atom.  Trust-exact can still converge to a very overfitted potential when the PBS is not balanced. BFGS does not overcome the ill-condition of the accurate Hessian Matrix in extreme cases. L-BFGS-B, originally developed to save the memory for the Hessian matrices and maintaining a lower-rank approximated Hessian matrix, has a better performance than BFGS. Trust region methods and Newton's method are more efficient than BFGS and L-BFGS-B as one can expect. In all the systems we tested, Trust-Krylov is the most robust method.\par
	\begin{table*}[t]
		\centering
		\caption{Optimizer Performance for the Ne Atom\textsuperscript{\emph{a}}}
		\begin{tabular}{|c||c|c|c|c|c|c|}
			
			\hline
			Method & $W[\Psi_{det}, v_{KS}]$ & $T_s[\Psi_{det}]$ & $\|grad\|$\textsuperscript{\emph{b}} & $N_{error}$ &$\epsilon_{\rm HOMO}$\textsuperscript{\emph{c}} & $N_f$\textsuperscript{\emph{d}}\\
			\hline
			Newton\textsuperscript{\emph{e}}+TSVD &128.609986 &128.597 &2.4e-2 &4.9e-3 &-0.6280 & 42 \\
			\hline
			BFGS &128.609810 &128.615 &1.6e-3 &8.1e-3 &-0.7452 & 12 \\
			\hline
			L-BFGS-B &128.609800 &128.614 &4.0e-3 &7.5e-3 &-0.7439 &10 \\
			\hline
			Trust-Exact &128.609948 &128.593 &1.5e-3 &5.2e-3 &-0.7010 &3 \\
			\hline
			Trust-Krylov &128.609926 &128.609 &6.6e-5 &3.0e-3 &-0.6988 &4 \\
			\hline
			Newton-CG & 128.609823 & 128.605 &2.2e-2 &4.2e-3 &-0.7408 &5 \\
			\hline
		\end{tabular}\\
		\textsuperscript{\emph{a}} Basis Set: CT/C5;
		\textsuperscript{\emph{b}} $L^2$;
		\textsuperscript{\emph{c}} The experimental I is 0.7925 \cite{west1979crc};\\
		\textsuperscript{\emph{d}} the number of function evaluations;
		\textsuperscript{\emph{e}} Strong Wolfe line search is used.\\
		PDE-CO: \textsuperscript{\emph{f}}3.5e-3; \textsuperscript{\emph{g}}-0.7460; \textsuperscript{\emph{h}}3.4e-3; \textsuperscript{\emph{i}}-0.7377.
		\label{OptMedTable}
	\end{table*}
	
	\section{Concluding Remarks}
	Pure density-to-potential iKS methods are in general more efficient but less accurate than wavefunction-based iKS methods. When implemented on finite PBS, with a further improvement on efficiency, the results are more sensitive. 
	Based on our experience, some of which is shown in the Appendix \textit{Example Calculations}, we recommend using the largest basis sets you can afford. Trust region methods, especially Trust-Krylov with $\lambda$-regularization, are usually more reliable (the search for $\lambda$ could start in a region $[10^{-4}, 10^{-6}]$). If a careful tuning is preferred, the Newton method with the null-space correction can be considered. Our code with all of the methods compared and discussed here will be open-sourced very soon, but a word of caution for the user is in order: pure iKS problems, especially with constrained optimizations, remain challenging. New ideas are needed.
	
	\begin{acknowledgements}
		This work was supported by the National Science Foundation under Grant No. CHE-1900301. The authors thank Cyrus Umrigar for providing the QMC data, and Qin Wu and Egor Ospadov for helpful discussions.
	\end{acknowledgements}
	
	\section{REFERENCES}
	\bibliography{ms}
	
\end{document}


\title{\bf Appendix: Example Calculations}
\date{}
\maketitle

We show the XC potentials for various systems with different methods:
\begin{itemize}
\item mRKS: the mRKS method from CISD wave functions.
\item WY: the Wu-Yang method with Trust-Krylov optimizer from CCSD densities.
\item PDE-CO: the PDE-constrained optimizations method with L-BFGS optimizer from CCSD densities.
\item WY+$\lambda$-reg: WY with $\lambda$-regularization.
\end{itemize}
All calculations below are performed on CD/CQ, which is a specifically chosen pair of basis sets that usually leads to unstable results as discussed in the main text. The results are plotted along the x-axis for 1D figure and xy-plane for 2D plots. Atomic units are used. Note that all XC potentials lead to convergence regarding the criteria discussed in the main article. We also report the $\epsilon_{\rm HOMO}$ in Table \ref{table:eHOMO}.\par
\begin{table}[H]
\centering
\caption{$\epsilon_{\rm HOMO}$ (atomic units)}
\begin{tabular}{|c||c|c|c|}
 \hline
system & WY & PDE-CO & WY+$\lambda$-reg\\
 \hline
 Ar &-0.5920 &-0.5926&-0.5737\\
 \hline
 H$_2$O &-0.3295 &-0.3290 &-0.3616\\
 \hline
 FC$_2$Cl &-0.3258&-0.2259&-0.3106\\
 \hline
 Ethanol &-0.2859 &-0.2953 &-0.2857\\
 \hline
 Benzene &-0.2654&-0.2742&-0.2679\\
  \hline
\end{tabular}
\label{table:eHOMO}
\end{table}

\begin{itemize}
\item
Ar\\
\begin{table}[H]
\centering
\begin{tabular}{l|l|l|l}
 \hline
&x&y&z\\
 \hline
Ar &0.0 &0.0 &0.0\\
 \hline
\end{tabular}
 \caption{Ar}
\label{list:Ar}
\end{table}
\begin{figure}[H]
        \centering
		\includegraphics[width=0.9\linewidth]{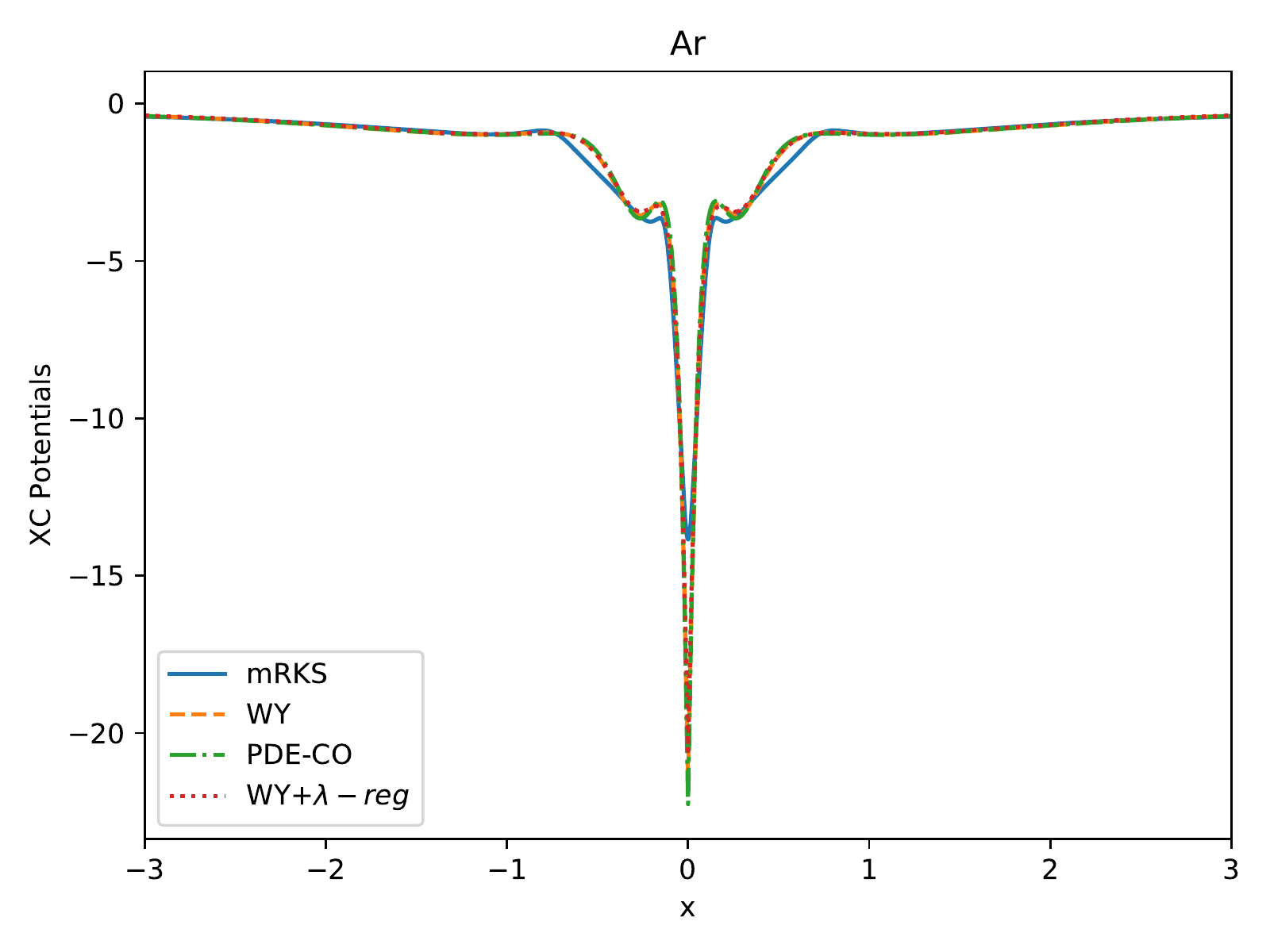}
		\label{Ar_CDCQ}
\end{figure}

\item

H$_2$O\\
\begin{table}[H]
\centering
\begin{tabular}{l|l|l|l}
 \hline
&x&y&z\\
 \hline
O  &0.0000& 0.0000& 0.0000\\
H &-0.4607& 1.8327& 0.0000\\
H  &1.8897& 0.0000& 0.0000\\
 \hline
\end{tabular}
 \caption{H$_2$O}
\label{list:H2O}
\end{table}

\begin{figure}[H]
        \centering
		\includegraphics[width=0.9\linewidth]{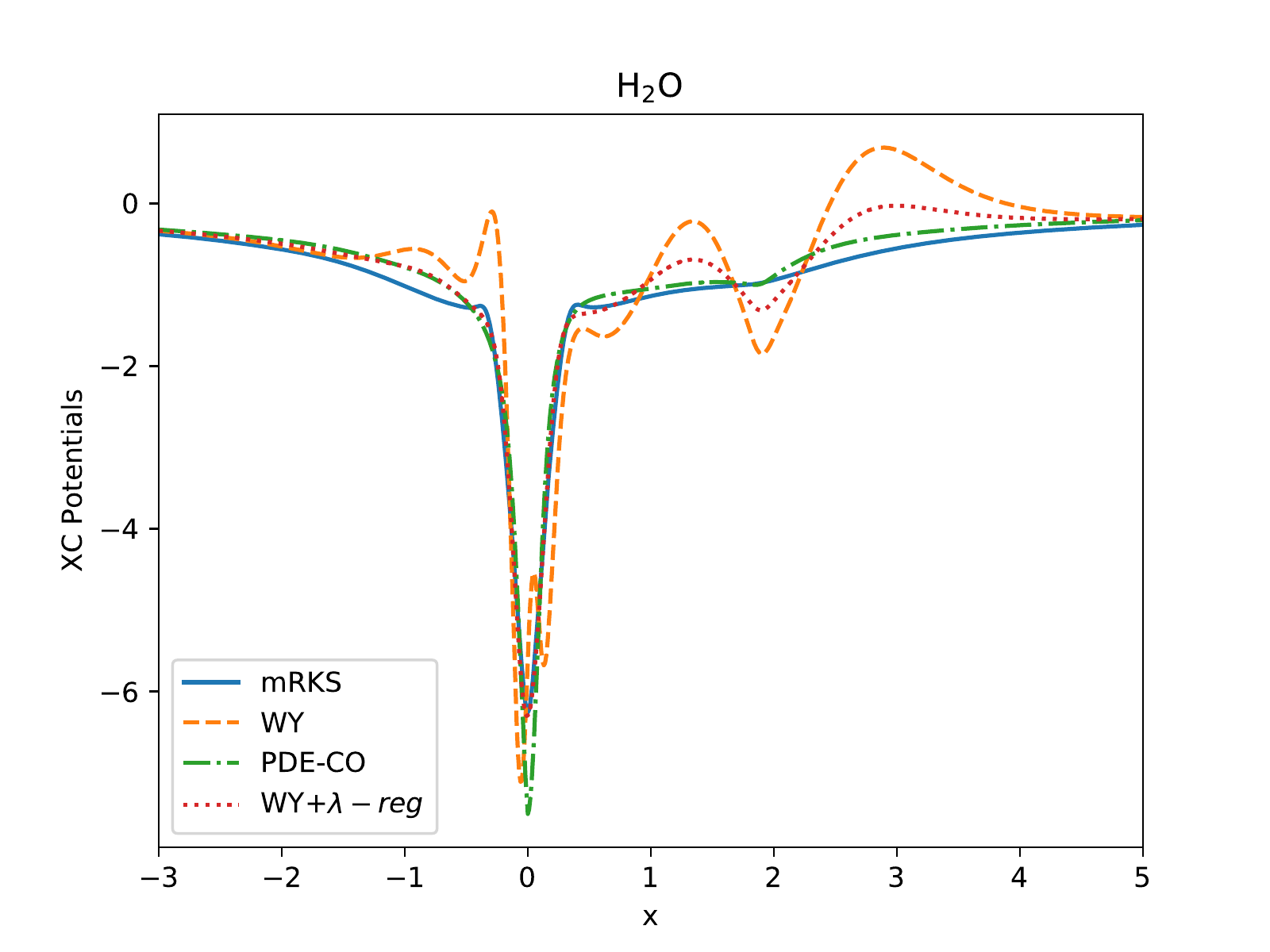}
		\label{H2O_CDCQ}
\end{figure}	
\item
FC$_2$Cl\\
\begin{table}[H]
\centering
\begin{tabular}{l|l|l|l}
 \hline
&x&y&z\\
 \hline
F &-4.6227 &0.0000 &0.0000\\
C &-2.2256 &0.0000 &0.0000\\
C &0.0000 &0.0000 &0.0000\\
Cl &3.1308 &0.0000 &0.0000\\
 \hline
\end{tabular}
 \caption{FC$_2$Cl}
\label{list:FC2Cl}
\end{table}

\begin{figure}[H]
        \centering
		\includegraphics[width=0.9\linewidth]{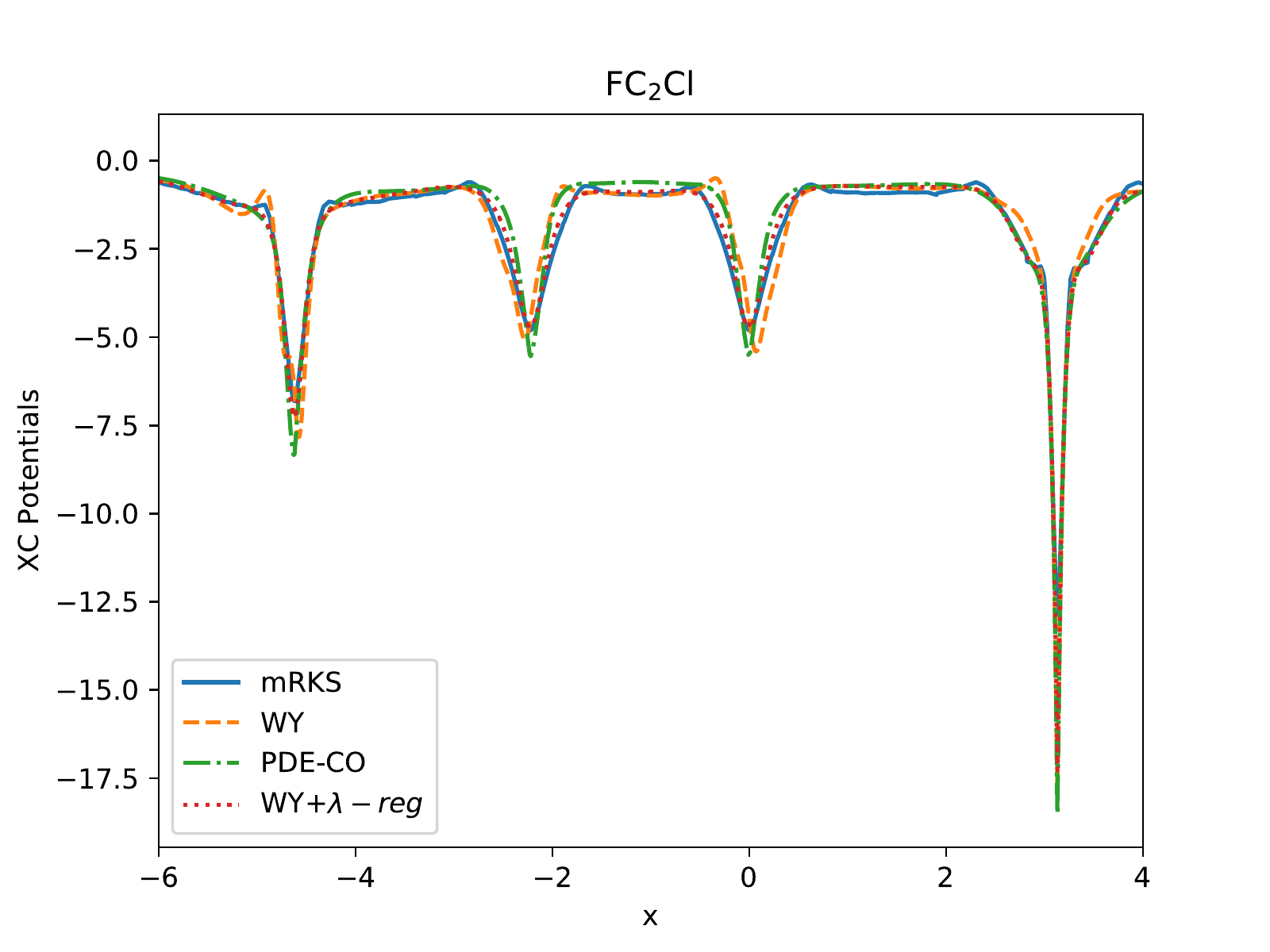}
		\label{FC2Cl_CDCQ}
\end{figure}

\item
Ethanol (C$_2$H$_5$OH)\\
\begin{table}[H]
\centering
\begin{tabular}{l|l|l|l}
 \hline
&x&y&z\\
 \hline
C &2.2448 & -0.7236 &  0.0000\\
C &0.0000 &  1.0443 &  0.0000\\
O &-2.2425 & -0.4671  &  0.0000\\
H &-3.6353 &  0.7275 &  0.0000\\
H &3.9656 &  0.4358 &  0.0000\\
H &2.1135 & -1.9073 &  1.6760\\
H &2.1135 & -1.9073 & -1.6760\\
H &-0.0429 &  2.2321  &  1.6729\\
H &-0.0429 &  2.2321  & -1.6729\\
 \hline
\end{tabular}
 \caption{Ethanol}
\label{list:Ethanol}
\end{table}

\begin{figure}[H]
        \centering
		\includegraphics[width=0.9\linewidth]{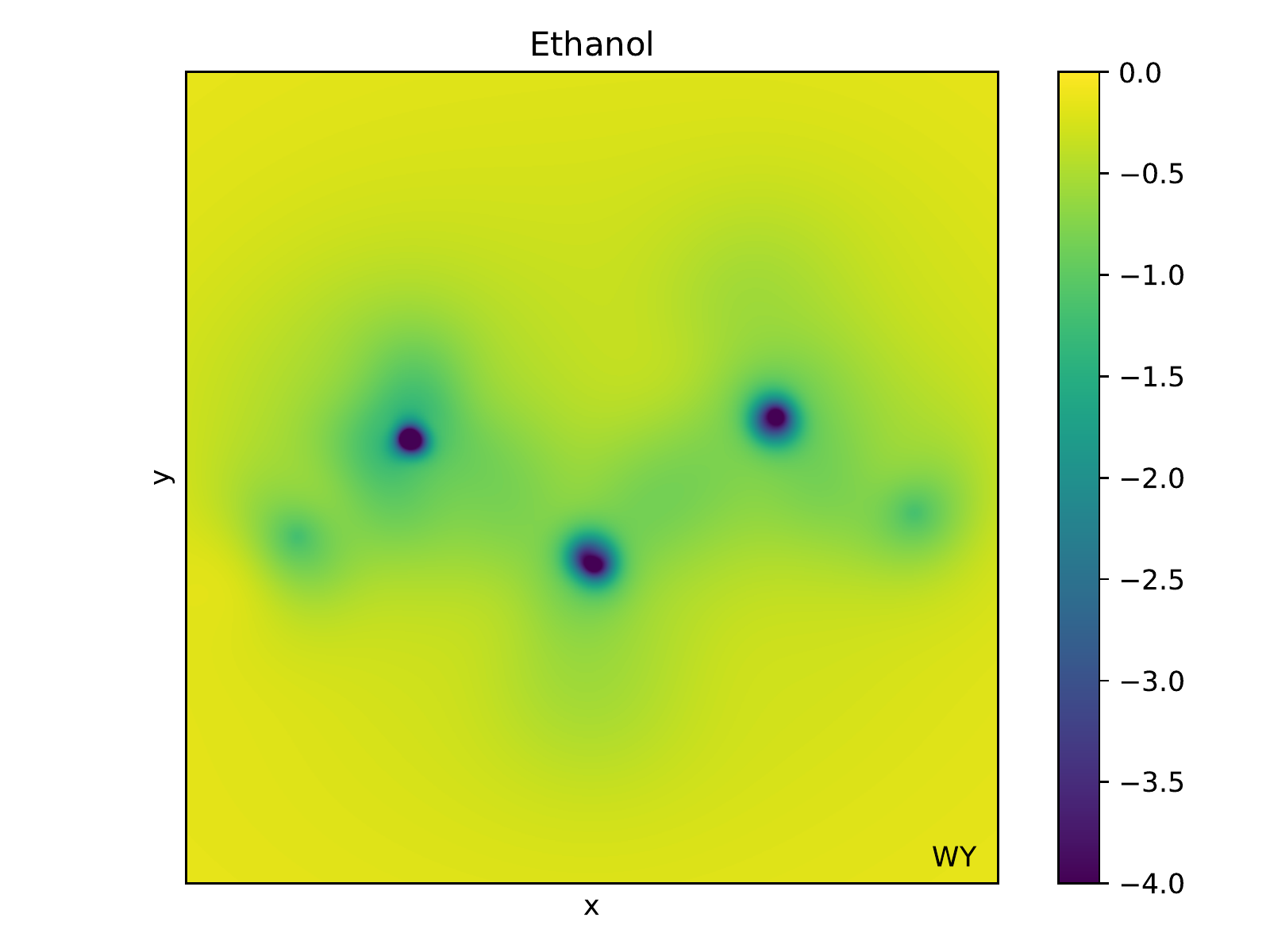}
		\label{Ethanol_CDCQ}
\end{figure}
\begin{figure}[H]
        \centering
		\includegraphics[width=0.9\linewidth]{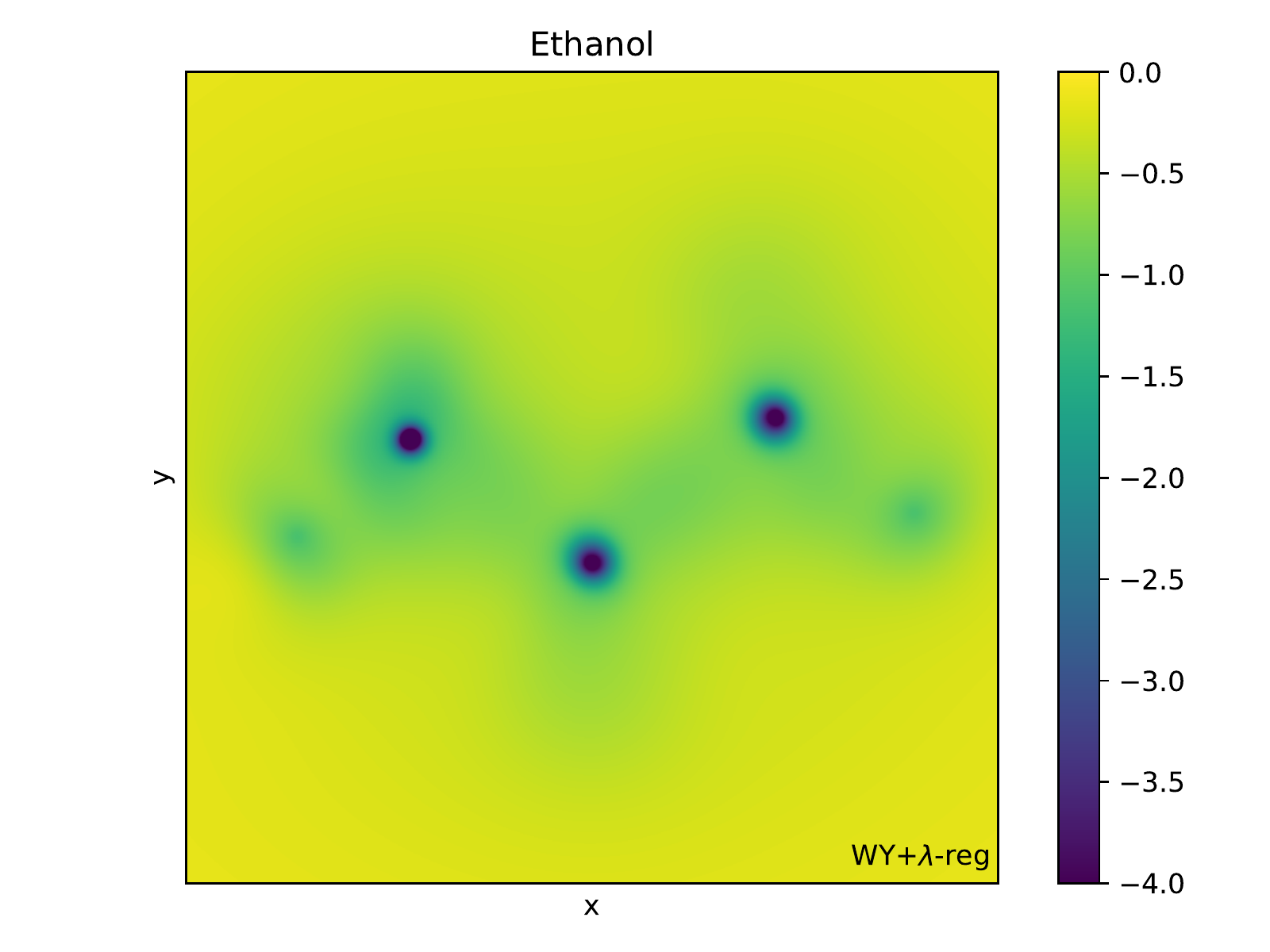}
		\label{Ethanol_CDCQ}
\end{figure}
\begin{figure}[H]
        \centering
		\includegraphics[width=0.9\linewidth]{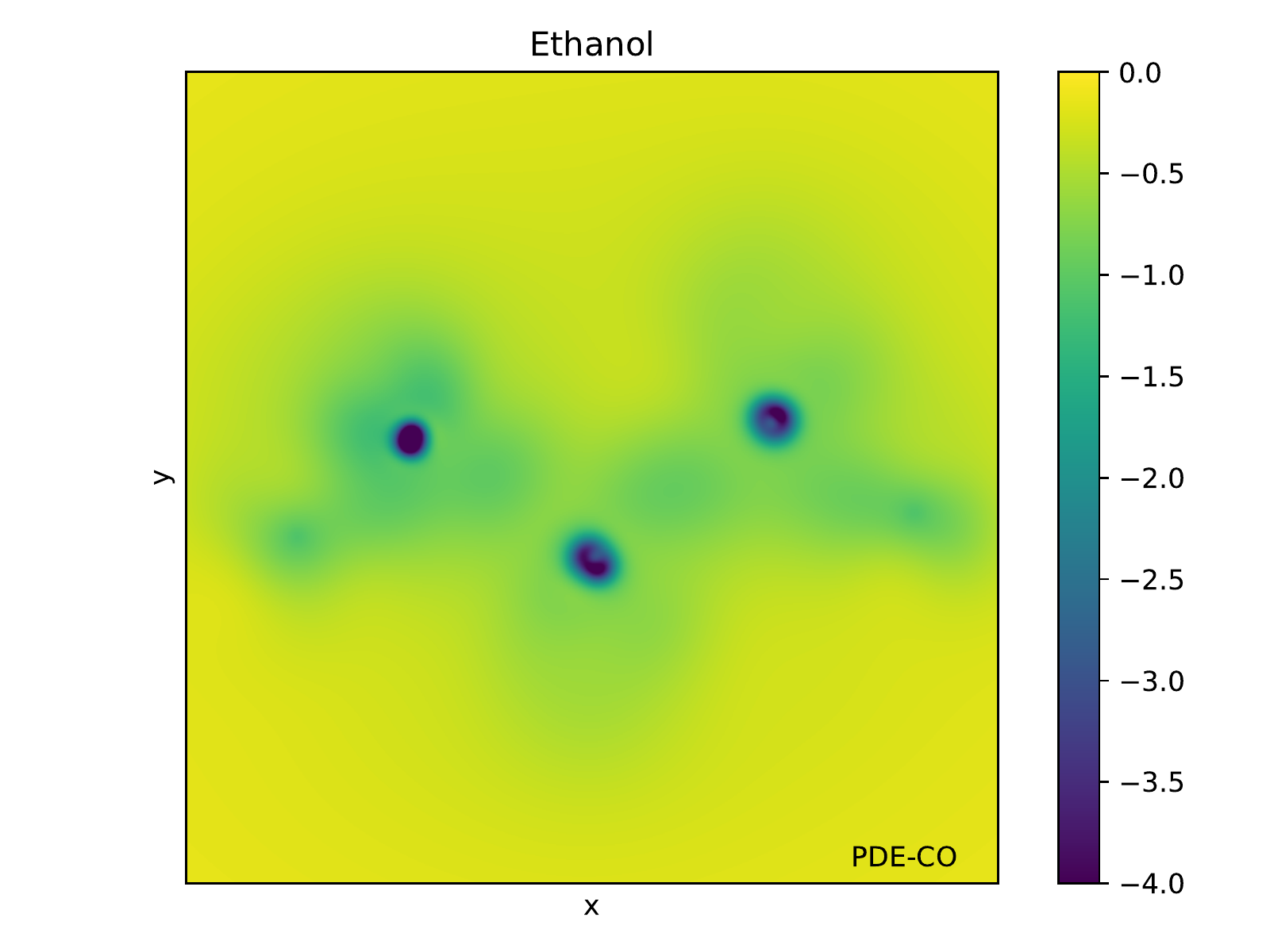}
		\label{Ethanol_CDCQ2}
\end{figure}

\item
Benzene\\
\begin{table}[H]
\centering
\begin{tabular}{l|l|l|l}
 \hline
&x&y&z\\
 \hline
C &-1.3105 & 2.2699 & 0.0000\\
H &-2.3329 & 4.0406 & 0.0000\\
C &1.3105 & 2.2699 & 0.0000\\
H &2.3329 & 4.0406 & 0.0000\\
C &2.6211 &  0.0000 & 0.0000\\
H &4.6657 &   0.0000 & 0.0000\\
C &1.3105 &  -2.2699 & 0.0000\\
H &2.3329 &   -4.0406 & 0.0000\\
C &-1.3105 &  -2.2699 & 0.0000\\
H &-2.3329 &  -4.0406 & 0.0000\\
C &-2.6211 &  0.0000 & 0.0000\\
H &-4.6657 &  0.0000 & 0.0000\\
 \hline
\end{tabular}
 \caption{Benzene}
\label{list:Benzene}
\end{table}

\begin{figure}[H]
        \centering
		\includegraphics[width=0.9\linewidth]{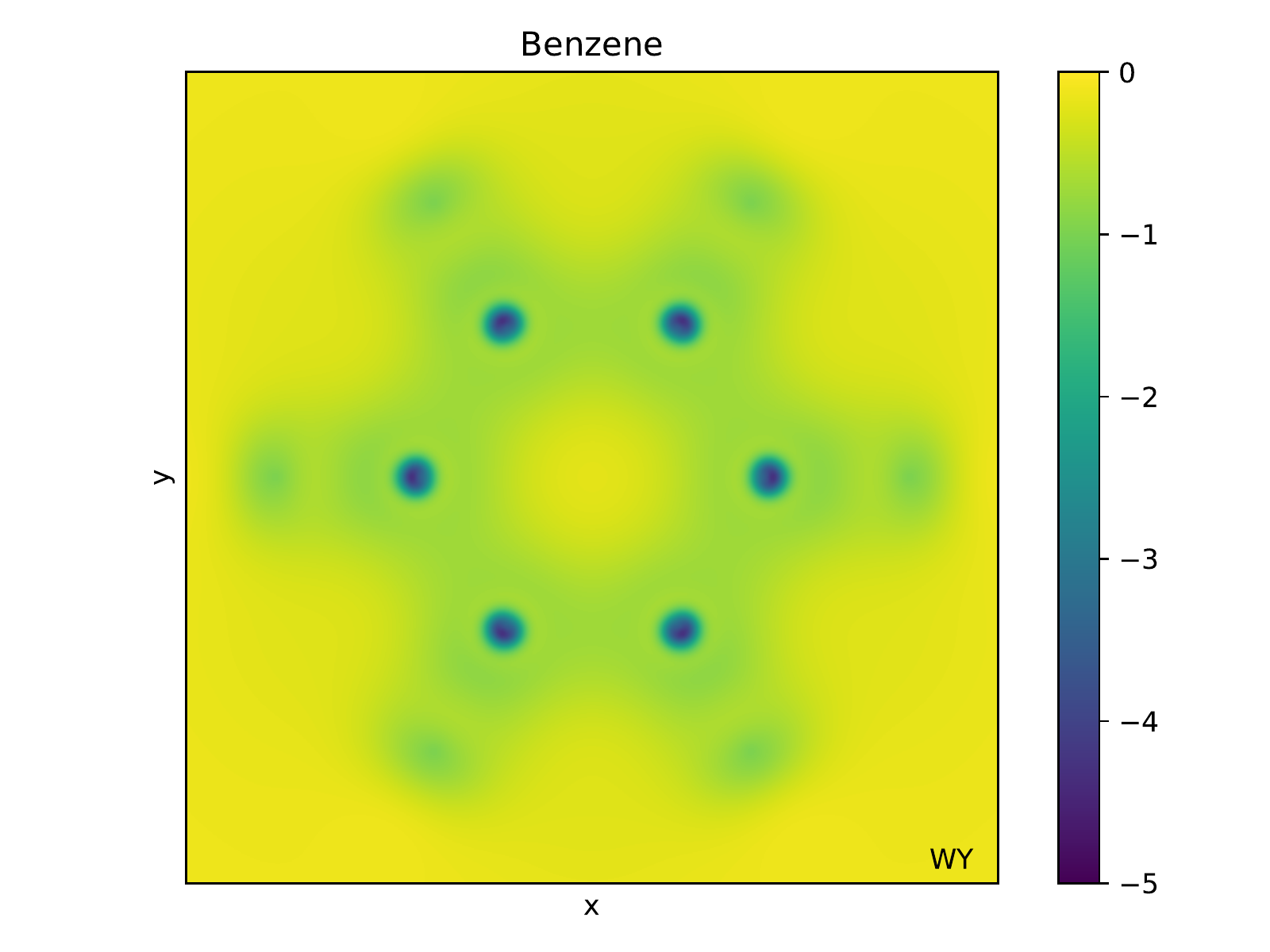}
		\label{Benzene_CDCQ}
\end{figure}
\begin{figure}[H]
        \centering
		\includegraphics[width=0.9\linewidth]{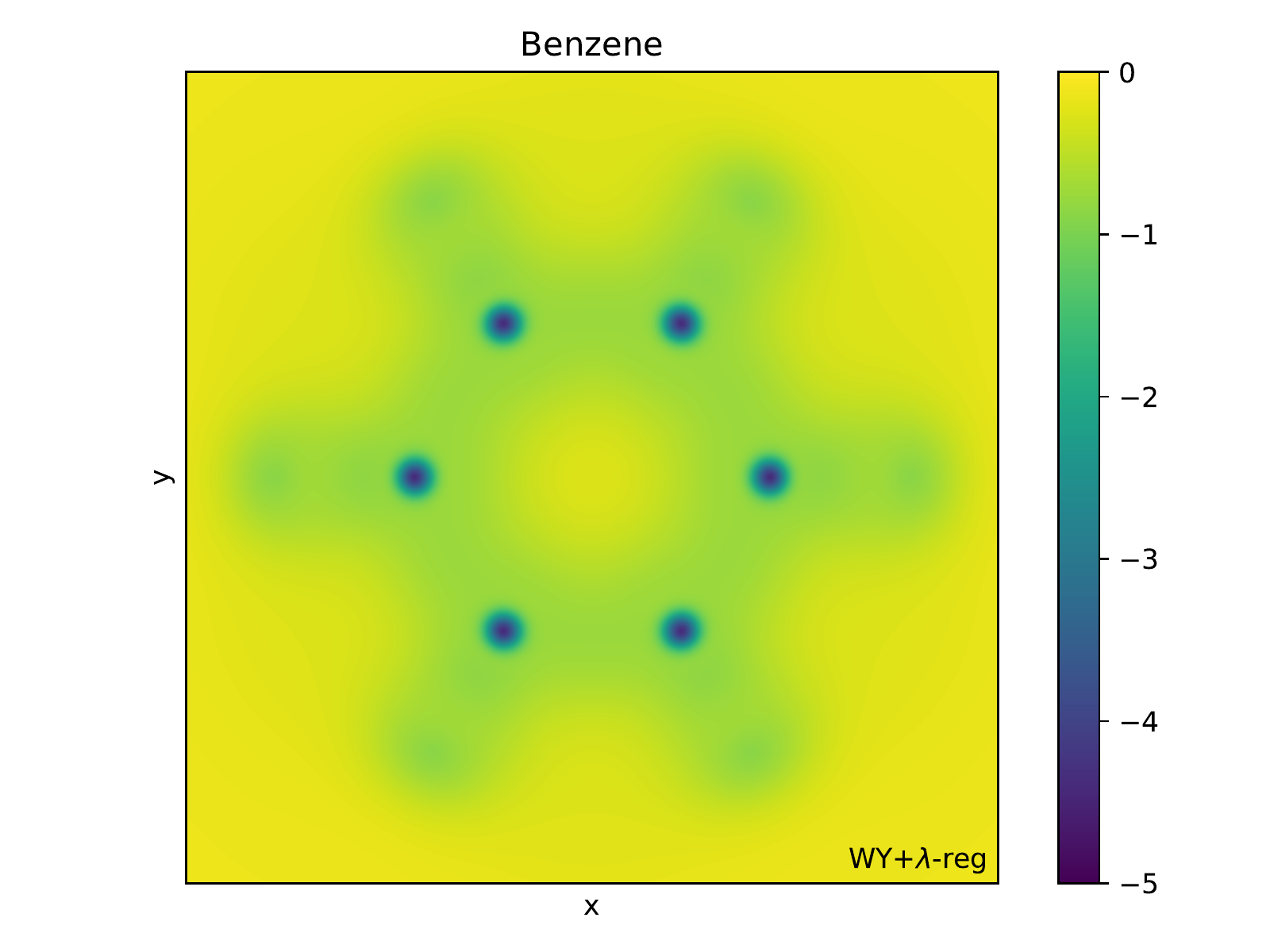}
		\label{Benzene_CDCQ}
\end{figure}
\begin{figure}[H]
        \centering
		\includegraphics[width=0.9\linewidth]{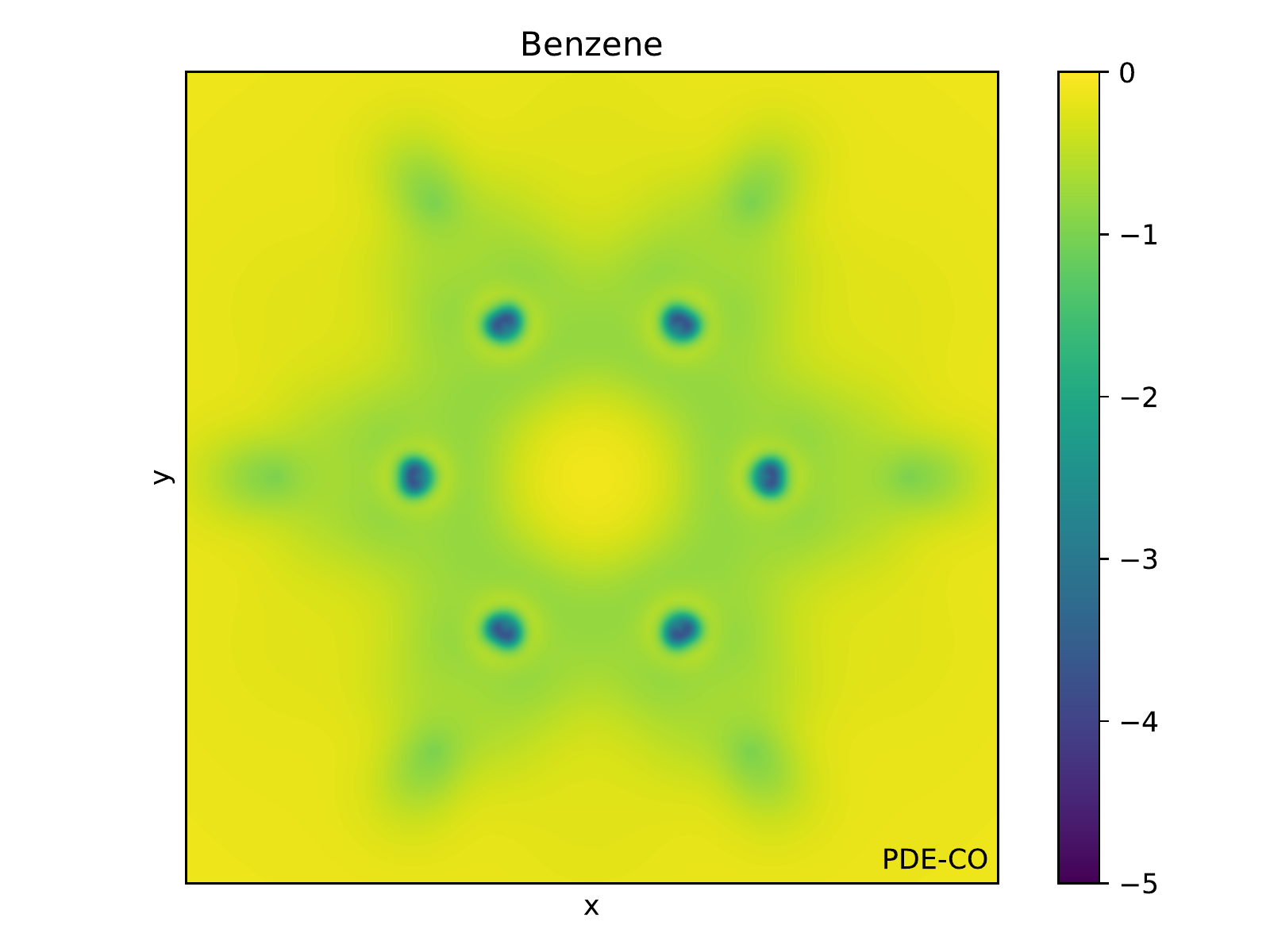}
		\label{Benzene_CDCQ2}
\end{figure}
\end{itemize}